\documentclass[sigconf]{acmart}

\AtBeginDocument{%
  }

\copyrightyear{2025} 
\acmYear{2025} 
\setcopyright{cc}
\setcctype{by}
\acmConference[CHI '25]{CHI Conference on Human Factors in Computing Systems}{April 26-May 1, 2025}{Yokohama, Japan}
\acmBooktitle{CHI Conference on Human Factors in Computing Systems (CHI '25), April 26-May 1, 2025, Yokohama, Japan}\acmDOI{10.1145/3706598.3714204}
\acmISBN{979-8-4007-1394-1/25/04}

\usepackage{longtable}
\usepackage{array}
\usepackage{makecell}
\usepackage{graphicx}
\usepackage{pdfpages}
\usepackage{subfig}
\usepackage{natbib}
\usepackage{longtable}
\usepackage{wrapfig}
\usepackage{siunitx}
\usepackage{multirow}

\usepackage{color, soul}

\begin{document}

\title[Towards Sensitivity and User Acceptance of Presence Questionnaires]{When Do We Feel Present in a Virtual Reality? Towards Sensitivity and User Acceptance of Presence Questionnaires}

\author{Annalisa Degenhard}
\affiliation{%
    \institution{Institute of Media Informatics, Ulm University}
  \city{Ulm}
  \country{Germany}
}
\email{annalisa.degenhard@uni-ulm.de}

\author{Ali Askari}
\orcid{0000-0002-4374-3635}
\affiliation{%
    \institution{Institute of Media Informatics, Ulm University}
  \city{Ulm}
  \country{Germany}
}
\email{ali.askari@uni-ulm.de}

\author{Michael Rietzler}
\affiliation{%
    \institution{Institute of Media Informatics, Ulm University}
  \city{Ulm}
  \country{Germany}
}
\email{michael.rietzler@uni-ulm.de}

\author{Enrico Rukzio}
\affiliation{%
    \institution{Institute of Media Informatics, Ulm University}
  \city{Ulm}
  \country{Germany}
}
\email{enrico.rukzio@uni-ulm.de}

\renewcommand{\shortauthors}{Degenhard et al.}

\begin{abstract}

Presence is an important and widely used metric to measure the quality of virtual reality (VR) applications. Given the multifaceted and subjective nature of presence, the most common measures for presence are questionnaires. But there is little research on their validity regarding specific presence dimensions and their responsiveness to differences in perception among users. We investigated four presence questionnaires (SUS, PQ, IPQ, Bouchard) on their responsiveness to intensity variations of known presence dimensions and asked users about their consistency with their experience. Therefore, we created five VR scenarios that were designed to emphasize a specific presence dimension. Our findings showed heterogeneous sensitivity of the questionnaires dependent on the different dimensions of presence. This highlights a context-specific suitability of presence questionnaires. The questionnaires' sensitivity was further stated as lower than actually perceived. Based on our findings, we offer guidance on selecting these questionnaires based on their suitability for particular use cases.\\
\end{abstract}

\ccsdesc[500]{Human-centered computing~Empirical studies in HCI}
\ccsdesc[500]{Human-centered computing~HCI theory, concepts and models}

\keywords{virtual reality, presence, plausibility, place illusion, embodiment, social presence, involvement, questionnaires}

\begin{teaserfigure}
  \includegraphics[width=\textwidth]{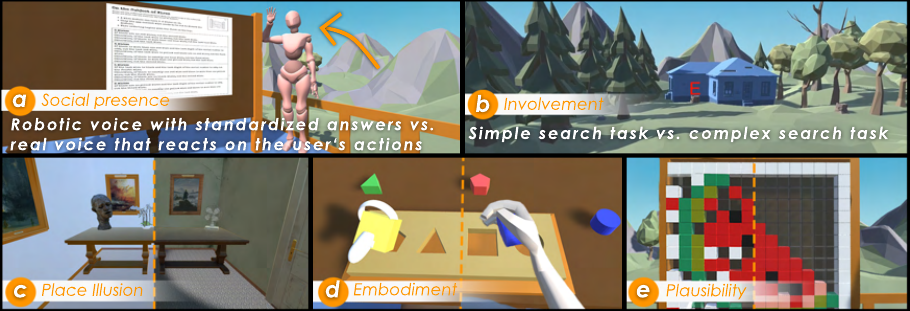}
  \caption{The five scenarios used in our study were designed to manipulate the intensity of certain dimensions of presence.}
  \Description{Teaser figure showing screenshots of our five scenarios implemented to manipulate certain dimensions of presence. On the upper left, the scenario for social presence is visualized, showing a low poly world with the avatar waving toward the observer. The observer is positioned in front of a pin board that shows the instructions for the scenario's task. A picture label explains that the task was performed once with the avatar having a real voice and responding to the user and once with a robotic voice. On the upper right, the scenario for the involvement dimension is illustrated, showing a low poly nature scenario with a house straight ahead of the observer. A stump is positioned in front of the house with a red letter hovering above it. The image label informs that the scenario included a simple or a complex search task. On the bottom left, two screenshots of the place illusion scenario with low detail on the left and high detail on the right are displayed. Both show the same room with a decorated table straight ahead, but the left image is shaded using simple textures, no light rendering, and shadow casting. The right picture of the two shows the room with detailed texture, light rendering, and shadow casting. A picture in the middle of the bottom shows two screenshots of the embodiment scenario. Both illustrate the same low poly scenario, with the observer looking at a table with a jigsaw puzzle on top. On the left, the hand of the observer is represented by a controller visualization holding a puzzle piece. On the right, the hand is represented by a low poly hand that holds another puzzle piece. On the bottom right, a screenshot is visualized that shows the scenario for the plausibility dimension. The observer is looking at a tiled board partly filled with colored cubes. The cubes indicate a pixeled image of a watermelon. On the left side, the image includes some errors in terms of incorrectly colored cubes. On the right, the image has no errors. The scenario's style is again low poly.}
  \label{fig:teaser}
\end{teaserfigure}

\received{20 February 2007}
\received[revised]{10 December 2024}

\maketitle

\section{Introduction}
The popularity of virtual reality has increased significantly over the last three decades. Allowing users to leave the real world and immerse into a virtual world, Virtual Reality (VR) has become popular not only in gaming but also in the industrial~\cite{hafsia_virtual_2018,squelch_ap_virtual_2001} and educational sectors~\cite{smutny_learning_2022}. Many studies observed a positive effect of VR on enjoyment, focus, and learning outcomes. VR allows the filtering of distracting factors and leads the user's focus on essential aspects. To be able to maximize the positive effect of fully immersing in the virtual world, research tried to identify what aspects of a VR experience influence the feeling of being there, which has been defined as \emph{presence}~\cite{held_telepresence_1992}. By clustering the influencing factors, \emph{dimensions of presence} were defined. The number of dimensions and the subjective character of presence make it difficult to measure it objectively. Therefore, questionnaires have become a common way to determine users' perceived presence. The most popular questionnaires regarding the number of citations in recent years~\cite{Hein2018TheUO} as well as in general~\cite{grassini_questionnaire_2020} are the SUS questionnaire~\cite{slater_virtual_2000,usoh_using_2000} by Slater, Usoh, and Steed, the IPQ (IGroup Presence Questionnaire) questionnaire~\cite{schubert_experience_2001,schubert_sense_2003} by Schubert and the PQ (Presence Questionnaire)~\cite{witmer_measuring_1994} by Witmer \& Singer. The questionnaires are widely used today to gain insight into the perceived presence of users in a virtual environment (VE). While the subjective nature of presence does not allow for a cross-study comparison of presence scores, insights on presence are often gathered by comparing different versions of a VR scenario and assigning differences in presence scores to the differences between the scenarios. However, presence questionnaires rarely include guidelines for their applicability to different VR solutions. In fact, they have been developed based on different understandings of presence and for different scenarios. Previous investigations have already drawn attention to the effects of questionnaire choice and the setting of a presence assessment on the outcome~\cite{schwind_using_2019,Hein2018TheUO,usoh_using_2000}. The context and wording, for example, may influence what a particular VR experience is compared to and, hence, influence the quality of a presence assessment. An assessment in reality is likely to bias respondents more towards comparing the VR experience to daily life experiences than an assessment in a virtual world would do. A lack of comprehension regarding the effects caused by design choices for a presence assessment may lead to misinterpretation and erroneous conclusions. With respect to that, we consider two limitations of the use of questionnaires in presence research:

The first one is that presence questionnaires are rarely analyzed for their \emph{sensitivity to certain dimensions of presence}. This makes it unclear if a questionnaire considers all dimensions of presence the researchers want to consider or that influences a user's experience respectively. Especially when it comes to comparative studies, as earlier mentioned, presence researchers would want to exclude the possibility that missing differences in presence scores might simply be caused by the chosen questionnaire not being sensitive to changes in the respective presence dimension. Considering the different underlying concepts of presence questionnaires, we suspect that the choice of the questionnaire could potentially bias the determination of the users' perceived presence by not being sensitive to some included presence dimensions of a virtual world. Single-item presence scales (SIP) like the one by Bouchard et al.~\cite{bouchard_reliability_2004} seem more neutral. However, in many cases, the item still contains a definition of presence, thus a certain bias on the user's part.

Second, it is rarely investigated whether \emph{respondents approve} to the scores of the questionnaires. Remembering the subjective nature of presence, not only being influenced by technical aspects but also by individual factors like suspension of disbelief \cite{vorderer_mec_2004} or motivation to get immersed \cite{klimmt_media_2003}, and considering the difficulty of assessing such an elusive phenomenon - not only for experienced presence researchers but especially for respondents who are, in many cases, not familiar with the concept - we think that it is important to consider not only sensitivity but also consistency of scores with the respondents' actual experiences.\\

Our goal was to investigate the ability of different presence questionnaires to assess changes in specific dimensions of presence. Therefore, we investigated (1) the \emph{sensitivity of presence questionnaires to known dimensions of presence} on the one hand and (2) the users' approval of the resulting scores considering their actual experience on the other hand. 

\paragraph{Sensitivity of Presence Questionnaires}
Since there are no guidelines on which presence questionnaire is best suited for a specific VR solution and on their individual strengths and weaknesses, it is still not clear which one to choose to determine users' presence with the highest validity. An adverse choice could cause a dimension of presence relevant to the respective VR solution not to be considered in the questionnaire. Therefore, results could lead to misinterpretations. 
We investigated the sensitivity of SUS, IPQ, PQ, and SIP to changes in the intensity of specific presence dimensions. To understand the potential bias by the choice of presence questionnaires for a VR solution, we wanted to answer three questions: 

\begin{enumerate}
    \item How sensitive are they to changes in specific dimensions of presence? (RQ1)
    \item Do users agree to the differences in scores affected by the changes of a specific dimension of presence? (RQ2)
    \item Can questionnaire items be clustered according to their sensitivity to changes in a specific dimension of presence? (RQ3)
\end{enumerate}

\noindent We picked five dimensions of presence based on two presence models by Skarbez from 2017~\cite{skarbez_survey_2017} and Hein from 2018~\cite{Hein2018TheUO}. Skarbez's model combines multiple presence models based on a systematic literature review of existing presence research (331 citations from 2018-2020). We additionally chose Hein's model because it considers further presence dimensions from popular presence models by Schubert (2001,~\cite{schubert_experience_2001}, 1075 citations from 2018-2022) and Slater (2003,~\cite{slater_note_2003} 619 citations from 2018-2022) that Skarbez chose not to include as stand-alone dimensions. This resulted in the dimensions of place illusion, plausibility, involvement, user embodiment, and social presence. 
To gain insight into the sensitivity of the four questionnaires to changes in one of the five dimensions, we designed a VR scenario for each dimension that emphasized aspects of that dimension while reducing aspects of other dimensions to a minimum. We then altered its intensity within the scenario, resulting in two conditions per dimension: one for the high-intensity scenario and one for the low-intensity scenario. We asked users to complete all questionnaires for each condition and compared the resulting scores.

\noindent Our findings indicate that the sensitivity of the questionnaires to changes in the presence dimension varies. The SUS and IPQ showed high sensitivity to changes in place illusion, whereas the PQ was most responsive to changes in plausibility. The SIP questionnaire showed the highest sensitivity across all dimensions but with a notable standard deviation in scores. None of the four questionnaires were particularly sensitive to the involvement, social presence, or embodiment dimensions. Our results suggest that researchers should carefully select a questionnaire based on the focus of their study to ensure sensitivity and validity. Thoughtful questionnaire choice is crucial when investigating specific aspects of presence to avoid misinterpretation.\\

\paragraph{Approval of Sensitivity}
Since the high sensitivity of questionnaires to changes must not imply consistency with the users' actual perceptions, we further investigated whether users agreed with the scores. Therefore, we asked participants whether the score differences between the two conditions of each scenario complied with their expectations. We found that, in many cases, participants would have expected greater changes. Only in the involvement and embodiment scenarios were the score differences of certain questionnaires higher than users expected. This not only underscores our findings in the previous research question that questionnaires may lack sensitivity when it comes to certain dimensions of presence but also confirms our assumption that greater sensitivity may not imply a realistic reflection of changes in scores. \\


Our evaluation pointed out the significant effects of the choice of presence questionnaires on the validity of presence assessments and provided various insights about the sensitivity of the questionnaires. The main findings of our investigations are:

\begin{enumerate}
    \item Findings on the sensitivity of presence questionnaires to different presence dimensions that should be considered for the choice of a questionnaire in the future.
    \item A validation of the questionnaires based on the subjective agreement of participants regarding their sensitivity to changes in individual presence dimensions.
    \item Findings about the sensitivity of individual questionnaire items to changes in different presence dimensions that should be considered when interpreting scores.
\end{enumerate}

\section{Related Work} 
Presence is the phenomenon of experiencing a virtual world and getting so involved in it that one starts to think that they are actually in the virtual world, often described as the "sense of being there"~\cite{lombard_at_1997}. Definitions of presence in research consider different aspects of presence, such as physical location, the sense of being elsewhere, their concurrent discrepancy, consciousness, awareness, or the users' actions. Common approaches mention the mental location of the user in the other place, meaning the virtual world~\cite{Metzinger2003BeingNO, lombard_at_1997}.  Slater et al., for example, defined presence in 1997 as "a state of consciousness, the (psychological) sense of being in the virtual environment"~\cite{slater_framework_1997}. Other definitions also focus on the aspect of consciousness but with regard to the awareness of the real world. 
Seth et al. defined presence as "the sense of now being in a VE while transiently unaware of one’s real location and the technology delivering the sensory input and recording the motor output."~\cite{seth_interoceptive_2012}. 
Witmer and Singer also based their approach on a user's perceived location but refined it by saying that the user has to believe what he is seeing to be able to be present. In 1994, they defined presence as the "subjective experience of being in one place when one is physically in another" and supplemented their definition by stating that "Presence is the result of a continuously updated interior model of the environment, stressing the necessity for suspension of disbelief"~\cite{witmer_measuring_1994}. This concept was one of the first to consider the perception of the VE. In 2004, Lee proposed a concept that focused entirely on the user's perception of the VE. He defined presence as "a psychological state in which virtual objects are experienced as actual objects in either sensory or non-sensory ways"~\cite{lee_presence_2004}. 
Sanchez and Flach provided a more functional perspective. They focused their definition on the users' ability to interact with a VE: “[\dots] the key to the approach is that the sense of “being there” in a VE is grounded on the ability to “do there”~\cite{sanchez2005}. Flach et al. considered a more general perspective on the functionality of a VE: "[\dots] the reality of experience is defined relative to functionality, rather than to appearances"~\cite{flach_1998}. 
Concepts of presence are of various nature. Providing a certain definition of presence is likely to influence the user's perspective on the phenomenon. Consequently, definitions in questionnaires are likely to bias the user's rating and should be taken into account when it comes to the choice of questionnaires.\\ 

\subsection{Presence Models}
Many aspects of VR experiences have been explored on possible correlations with presence~\cite{steuer_defining_1992, welch1996, gorini_role_2011, witmer1998measuring, skadberg2004visitors,roth_2016_social,lombard_at_1997}. While factors of immersion seem to influence presence in most VR experiences, it is still unclear which user-specific aspects influence presence and to what extent. For some, auditory cues might be the most relevant aspects of virtual worlds to be able to feel present. For others, it may be visual cues. The facets of an individual that can influence a VR experience are various and may even vary per individual and experience. For this reason, it is difficult to provide an exact model of presence. Different approaches have been published over the years. We will use the term \emph{dimensions of presence} in the following to describe clusters of aspects that are assumed to influence presence according to previous work.\\ 

\textit{Historical Foundations of Presence Models}\\
Sheridan et al. (1992) proposed one of the first models. They identified five dimensions that support the feeling of telepresence: the extent of sensory information, the control of sensors relative to the VE, the ability to modify the environment, task difficulty and the degree of automation. Since telepresence was defined to describe the phenomenon of presence in the context of workstations~\cite{sheridan_musings_1992}, Sheridan et al. focused their approach primarily on factors of immersion. However, the concept has similarities with models of presence in general. Zeltzer proposed a comparable model in 1992~\cite{zeltzer_autonomy_1992}. For him, the quality of a simulation system is influenced by the degree of a user's autonomy, the quality of interaction, and presence.\\
\noindent Steuer (1992) was the first to differentiate between technical aspects and the resulting subjective experience~\cite{steuer_defining_1992}. According to Steuer, the ability to perceive telepresence is influenced by the quality of sensory feedback (vividness) and interaction (interactivity). They split up the concept of telepresence into immersion as the technical aspect of a VE and presence as the subjective experience of an immersive system. Since then, several models of presence dimensions have been proposed that focus more on the user experience of a VE~\cite{slater_framework_1997}. Lombard and Ditton categorized five dimensions of presence: spatial presence, sensory presence, social realism, social presence, and engagement~\cite{lombard_at_1997}. Similarly, Heeter et al.~\cite{heeter_being_1992} stated that presence consists of personal presence, social presence, and environmental presence.\\

\textit{Psychological Models} With the publication of more perceptual perspectives on presence, psychological approaches began to appear. One was published by Waterworth et al. in 2001~\cite{waterworth_focus_2001}. They considered presence to be influenced by Locus (whether a user's attention lies on the virtual or real world), Focus (the user's state on a continuum scale from being absent to being present), and Sensus (the user's arousal on a continuum scale from unconscious to conscious). Their concept shows similarities to Lombard and Ditton's engagement but focuses entirely on perceptual involvement with the VE.\\

\textit{Differentiation between Place Illusion and Plausibility} A frequently used model in the last years has been proposed by Slater, who extended his concept of presence~\cite{slater_framework_1997} by introducing place illusion (PI) and plausibility (Psi) and by noting the relevance of the illusion of body ownership~\cite{slater_place_2009, slater_separate_2022}. He criticized the lack of specification of the "there" in the definitions of presence and the consideration of the individual differences of users. PI is defined as "the illusion of being at the place depicted by the VR". Orthogonal to PI, Psi describes "the illusion that the virtual situations and events are really happening"~\cite{slater_separate_2022}. Comparable to plausibility, Weber et al. proposed a presence model which states that presence occurs "if a mediated environment captures and maintains the attention of users and is perceived as realistic ~\cite{weber_how_2021}. The difference is that realism is "not just an illusion but rather the result of a conscious evaluation of the credibility and realness of the VE". Lee et al.~\cite{lee_presence_2004} considered presence to be composed of social presence, physical presence, and self presence. In this context, he defined social presence as "the sense of being together with another and mental models of other intelligences (i.e., people, animals, agents, gods, etc.) that help us simulate other minds", physical presence as "a psychological state in which virtual (para-authentic or artificial) physical objects are experienced as actual physical objects in either sensory or non-sensory ways." and self presence as "a psychological state in which virtual (para-authentic or artificial) self/selves are experienced as the actual self in either sensory or non-sensory ways". Again, physical presence can be compared with place illusion and self-presence with body ownership.\\

\begin{figure}[t]
    \centering\includegraphics[width=\columnwidth]{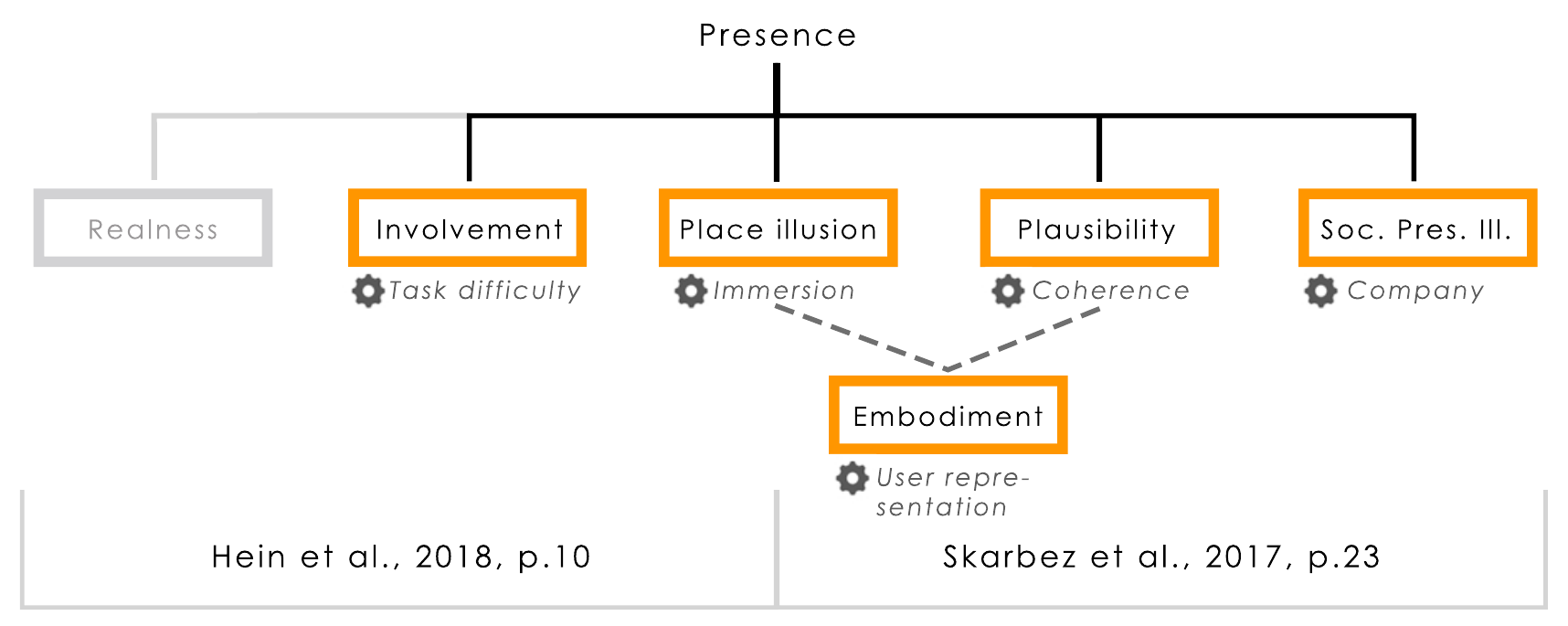}
    \caption{Overview of the presence dimensions that we chose for our investigations split by the two presence models considered. On the right: Dimensions of the presence model by Skarbez et al. (2017, p.23), which summarizes previous models based on a literature review~\cite{skarbez_survey_2017}. On the left: Dimensions of the presence model by Hein et al. (2018, p.10), which summarizes the presence models of Slater and Schubert et al.~\cite{Hein2018TheUO}.}
    \Description{Model for presence the project was based on. Presented as a hierarchical diagram. In the middle on the top is presence. Five dimensions, realness, involvement, place illusion, plausibility, and social presence, are all listed at the same level below presence, all directly connected to presence. Below PI and Psi, the dimension embodiment is positioned and connected via dotted lines to both of them. Below each dimension, the factor of the respective dimension is stated, except for realness, which is grayed out, as it won't be considered in this work. A line at the bottom of the model separates the named dimensions into two parts. The left side is labeled with Hein et al. and includes realness, involvement, and place illusion. The right part, labeled Skarbez et al. also includes place illusion, as well as plausibility, embodiment, and social presence.} 
    \label{fig:presence_models}
\end{figure}

\textit{Combined Presence Models} In their 2017 literature review, Skarbez et al. identified further similarities between prominent models of presence~\cite{skarbez_survey_2017} and derived their own model. As this model combines multiple concepts of presence, we decided to consider it for our investigations. The dimensions considered from Skarbez's model can be seen in~\autoref{fig:presence_models} on the right. However, Skarbez's model does not consider attentional aspects that have been part of several presence models~\cite{witmer_measuring_1994,lessiter_cross-media_2001,schubert_experience_2001}. Skarbez notes that the term is used differently in the literature and needs to be defined more precisely. However, since we consider attentional resources to influence presence, we decided to include a second model of presence further.\\

\noindent Hein et al.~\cite{Hein2018TheUO} proposed a model for presence that combined Slater's distinction between immersion and presence~\cite{Slater1997AFF} with Schubert et al.'s model. Based on the work from Witmer and Singer in 1998 and 1999~\cite{witmer1998measuring, witmer_measuring_1999}, Slater considered presence to result from immersion and user characteristics~\cite{slater_note_2003}. 
Based on this concept, Schubert, Regenbrecht, and Friedman (2001) conducted a factor analysis. They used items from several presence questionnaires and three first-order presence factors, including involvement, spatial presence, and realness~\cite{schubert_experience_2001}.
Involvement describes the user's focus during the mediated experience. Whether they focus more on the virtual or real environment and how captivated they are by the virtual environment.
The factor of spatial presence is about the actual sense of being there. The factor loads high when the users perceive the mediated environment as the actual environment in which they are located and with which they can interact. 
The realness factor summarizes the consistency of events in the virtual environment with real-world experiences and the extent to which the virtual environment is perceived as real.
Hein et al. combined these results with Slater's approach to a presence model. The considered presence dimensions of this model can be seen in \autoref{fig:presence_models} on the left.\\ 

\noindent\textit{Comparison of the Presence Models} The models by Skarbez and Schubert share multiple similarities. Slater's place illusion by Slater can be compared with the concept of spatial presence by Schubert. Both concepts describe the illusion of actually being in the environment depicted by the media system. Plausibility and realness also share similarities, but plausibility is more precise. While realness describes the consistency with real-world experiences in general, which leaves room for interpretation, plausibility is defined as the illusion that the depicted scenario is actually occurring. Slater focuses his concept on the believability of events happening in the virtual scenario, opening up the stiff concept of realness to experiences that may not be realistic but are reasonable. Slater states that as long as the users are able to construct a mental model of the virtual environment, they will be able to experience presence. Involvement and social presence may have similarities, but these concepts should be considered distinct. One can perceive a virtual avatar as a social companion in the VE but not interact with it and thus not necessarily be involved. On the other hand, one can be involved in another task in the VE while being alone. Thus, the user may be involved but not necessarily have the illusion of social presence. Both concepts state the influence of immersion on presence. Schubert further explicitly considers user characteristics as an explicit factor. This differs from Skarbez’s concept. It is not explicitly considered, but user characteristics influence the transitions from immersion, coherence, and company to the subsequent concepts. Individual differences will influence whether and how one experiences place illusion when exposed to a certain level of immersion.\\

\textit{Choice of presence models for this study:} We decided to consider both models of presence in our study. Both are based on prominent models of presence, and we were interested in gaining insight into both involvement and social presence. Therefore, we chose the following dimensions of presence: place illusion, plausibility, social presence, user embodiment, and involvement. We used place illusion and plausibility as defined by Slater et al.~\cite{slater_place_2009}. To investigate the effects of changes in place illusion, we manipulated immersion as a direct factor according to the model by Skarbez et al.~\cite{skarbez_survey_2017}. Regarding plausibility, we focused on the consistency of events in the VE with expectations. We thus decided to manipulate coherence. Events that were inconsistent with expectations would likely make the user question their plausibility, while expected events should improve plausibility. We decided to further investigate Slater's concept of body ownership as a separate dimension. As previously mentioned, Slater does not consider it a separate dimension but rather a component of place illusion and plausibility. However, as the direct influence of a user representation on presence has been shown by many researchers, e.g.~\cite{zhang_influence_2020,pan_foot_2019,bartl_avatar_2022,costa_embodiment_2013}, we expected that this separation would provide additional interesting insights. We will refer to this dimension as user embodiment. To investigate the effects of changes in user embodiment, we altered the representation of the user in the VE. We distinguished between social presence and involvement to investigate the effects of demanding tasks and the presence of other individuals on the different presence scores. Since the questionnaires chosen for this work treat involvement and social presence differently, we expected that this distinction would provide more meaningful insights.\\

\subsection{Presence Questionnaires} 
 Many approaches to measuring presence have been introduced over the years. Due to the scope of our project, we decided to limit our examination to some of the most significant surveys in terms of their frequency of use. The questionnaires considered are listed in the table~\ref{tab:questionnaires_overview}. We selected the following four presence questionnaires:
 \begin{enumerate}
    \item\emph{The Slater-Usoh-Steed questionnaire (SUS)} by Slater, Usoh, and Steed (2000) measures presence with six items. Unlike other presence questionnaires, the \emph{SUS} questionnaire focuses on the subjective experience rather than specific aspects of presence~\cite{slater_virtual_2000, usoh_using_2000}. Although published in 2000, the SUS is still one of the most frequently used presence questionnaires.

    \item\emph{The Presence Questionnaire (PQ)} by Witmer and Singer (1998)~\cite{witmer1998measuring} consists of 19 items with four subscales: involvement/control, sensory fidelity, naturalness, and interface quality. According to Hein et al., it is still by far the most widely used questionnaire in terms of the number of citations.

    \item\emph{The IGroup Presence Questionnaire (IPQ)} developed by Schubert (2001) consists of 14 questions with subscales on spatial presence, involvement and realness~\cite{schubert_experience_2001, schubert_sense_2003}. The IPQ focuses less on the design aspects of the VR system and the VE but rather on the user's perception of the VE.

    \item\emph{The single-item presence scale (\emph{SIP})} created by Bouchard (2004)~\cite{bouchard_reliability_2004} was also included in our research. The single-item approach may help to isolate the effects of the VR content itself on presence perceptions, as any variation in presence ratings can be more confidently attributed to the content rather than to variations in the measurement tool.
 \end{enumerate}

 \noindent Our decision was based on literature reviews on the use of presence questionnaires in the years 2016-2017~\cite{Hein2018TheUO} and 2002-2019~\cite{grassini_questionnaire_2020}. In addition, we aimed to use questionnaires with distinct dimensions to identify potential advantages in different use scenarios.\\
 To summarize, we decided to use a combination of Skarbez's presence model~\cite{skarbez_survey_2017} and by Hein et al.'s~\cite{Hein2018TheUO} (\autoref{fig:presence_models}). This resulted in a total of five dimensions, hereafter referred to as the \emph{place illusion} dimension (PI), the \emph{plausibility} dimension (PSI), the \emph{involvement} dimension (INV), the \emph{social presence} dimension (SOC), and the \emph{user embodiment} dimension (EMB) in the following. Our goal was to evaluate the sensitivity of our four picked questionnaires for these dimensions. Some of the questionnaires' underlying concepts consider parts of the dimensions, and some are not considered by any of the questionnaires, such as social presence. If we find differences in sensitivity and user approval, this would mean that the choice of the questionnaire would introduce a bias to presence assessments and that presence research should try to raise awareness for that. More refined guidelines for the choice of questionnaires would be needed to support a more user-centered and case-sensitive choice of questionnaires and promote the validity of presence assessments. 
 
 Therefore, we designed methods to investigate isolated effects of changes in a specific presence dimension and for the assessment of users' approval. The former is challenging as changes in assessed scores must be attributable to the considered dimension to derive meaningful insights for dimension-specific sensitivity. The latter has the difficulty that it should be avoided to ask users to rate absolute presence scores as the subjective phenomenon is difficult to quantify. The following section explains our approach to assessing dimension-specific sensitivity and user approval concerning the challenges provided by the subjective construct.

\section{Study Design}
The study aimed to evaluate the sensitivity of common presence questionnaires to changes in a specific dimension and how well this corresponds to the actual user experience. The selection of the four questionnaires resulted in a total of 40 questions (see Appendix: SUS:~\autoref{tab:questionnaire_sus}, PQ:~\autoref{tab:questionnaire_pq}, IPQ:~\autoref{tab:questionnaire_ipq}, SIP:~\autoref{tab:questionnaire_bouchard}). We did not examine the sensitivity of the subscales of the IPQ because we wanted to focus our work on measuring the overall concept of presence. To gain insight into how closely participants agreed with the results of particular questionnaires, we would have had to introduce additional concepts to the participants to help them understand the terminology. This would have affected the participants' ratings of the overall presence scores, as we would have introduced a bias for certain aspects of presence. These five dimensions and four presence questionnaires formed the basis of our study design to answer our research questions. \\

\subsection{Sensitivity of Presence Questionnaires} 
Our aim was to gain insight into the relevance of the four questionnaires for different use cases in VR. To achieve this, we designed scenarios that allowed us to isolate one of the five chosen dimensions at a time. This allowed us to understand the relationships between changes in a particular presence dimension and changes in overall presence scores. In designing the changes in each scenario, we had to take into account that the boundaries of dimensions for presence are unclear, and modifications to certain aspects of one dimension could also affect others. Therefore, we designed the scenarios to focus the experience on a single dimension with as little impact on the other dimensions as possible.
We developed two versions of each scenario - one in which the intensity of the dimension is reduced to a minimum, referred to as the \emph{negative} condition, and another with a high intensity, referred to as the \emph{positive} condition in the following. Minimum, in this case, refers to the reduction of these dimensions to an intensity that is necessary to prevent participants from being distracted by their absence. Aspects of the experience related to dimensions that are not in focus should remain unchanged between conditions. The detailed design decisions for each dimension and scenario are described below.

\begin{figure}[t]
        \centering
            \includegraphics[width=\columnwidth]{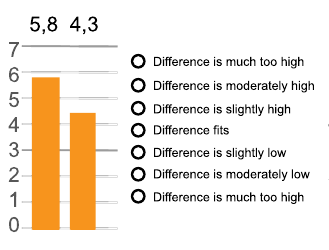}
            \caption{Example of a paired bar chart showing the scores of positive and negative condition. The score of each bar was shown above the bar. A radio button list on the right allowed users to state their approval.}
            \Description{Example of paired bar chart used to assess users’ approval. The chart has horizontal lines in the background that are labeled with numbers on the left that indicate scores from 0-7. The left bar ranges from the bottom to almost the seventh line. The right bar ranges from the bottom to slightly above the fifth line. Above each bar, a score the bars score is also written textually. On the right of the bar chart, a textual list with radio buttons before each list item is illustrated. The items are the ones described in the text ordered from "Difference is much too high" at the top to "Difference is much too low" at the bottom.}
           \label{fig:presence_scoredifferences}
\end{figure}

\subsection{Approval of Sensitivity}
As mentioned earlier, we also wanted to gain insight into the participants' agreement with the presence scores obtained from the various questionnaires. The difficulty with this investigation was that due to the subjective nature of presence, which is influenced by a variety of factors, it can be anticipated that participants will have difficulty quantifying an exact value of the difference in presence that they actually perceived. Consequently, we asked participants to rate the differences in scores on all four questionnaires after each scenario with both conditions. The scores of the negative and positive conditions were presented as bars in a paired bar chart (\autoref{fig:presence_scoredifferences}). To counterbalance any effects, the order of the four paired bar charts was randomized, and participants were not told which questionnaire each chart corresponded to. They were then asked to rate how much the difference between the scores corresponded to their actual perception of the differences between the conditions. Therefore, they were asked to rate the difference on a 7-point Likert scale ranging from -3 ="difference is much too low" to 3 = "difference is much too high", with 0 indicating that the difference corresponded to the participant's actual perception.

\subsection{Scenario Design}
    \begin{figure}[t]
        \centering
            \includegraphics[width=\linewidth]{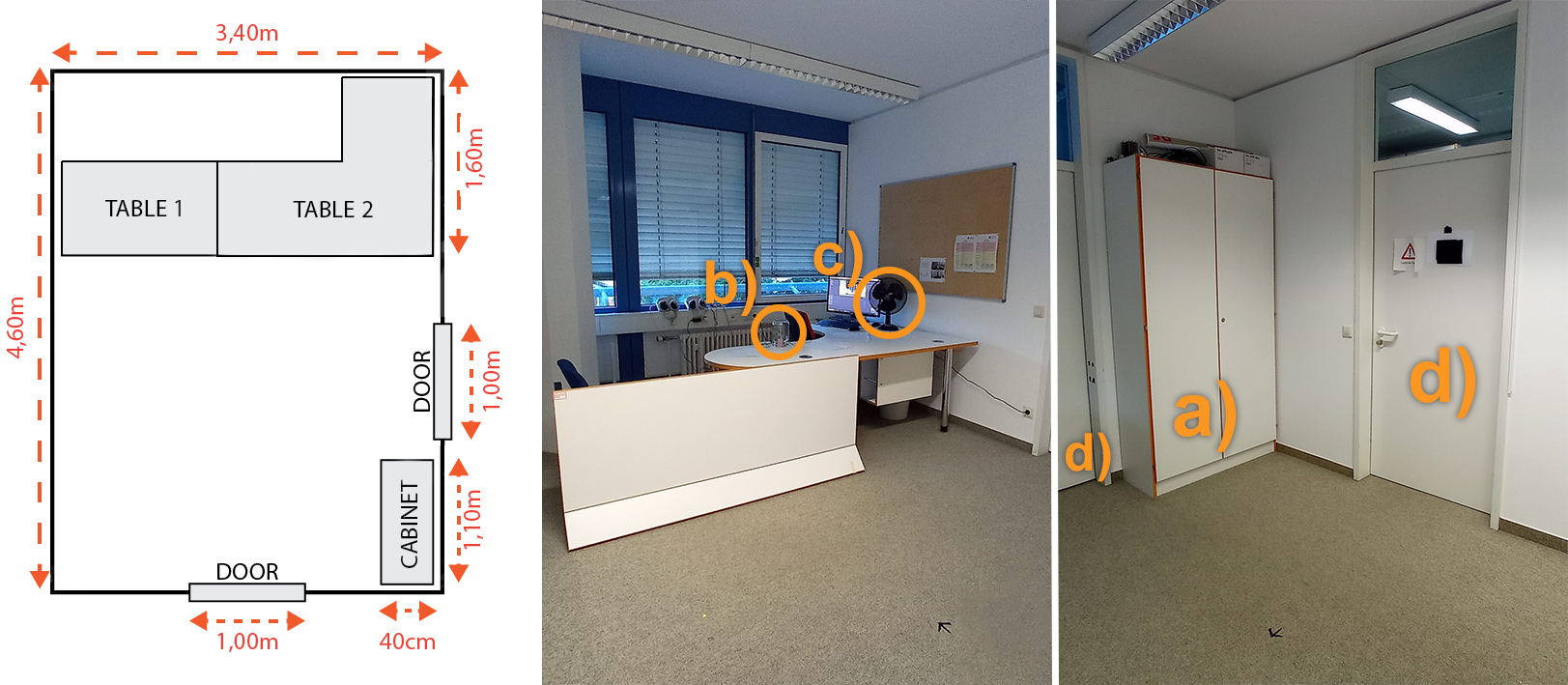} 
            \caption{Left: Dimension of the real room and the object placed in it. Middle \& right: Photos of the study room and the interaction space from the right back corner and the right front corner of the room. a) The cabinet that is recreated in the virtual scenario, b) The scented candle that creates the olfactory stimuli, c) The fan that creates the haptic stimuli, d) The doors that are placed in the same position in the virtual scenario.}
            \Description{Comparison of study room with virtual scenario. The figure is three parted. Left is a room plan for the study room from a top-down perspective. Labels inform about the room's dimensions and important distances. The plan also illustrates the positions of doors and furniture. The middle and right images are photos of this room. The photo in the middle shows the room with perspective on the windows and the tables that were positioned in front of them. Orange letters highlight the positions of the fan and olfactory candle on the table.}
            \label{fig:study_setup}
        \end{figure} 

We investigated a questionnaire's dimension-specific sensitivity and users ' approval using these two approaches, tailored to the difficult, elusive nature of presence. We restricted our evaluation to the introduced five dimensions and four questionnaires, but in theory, the approach can be applied to any other constellation as well.\\

The study was conducted in a room measuring ${\qty{3.40}{\metre} \times \qty{4.60}{\metre}}$, divided into a control area for the supervisor, separated by two tables from the study area where the participants moved around. The study area was approximately ${\qty{2.65}{\metre} \times \qty{3.00}{\metre}}$ and had a cabinet and two doors at the back and right side of the room (see~\autoref{fig:study_setup}). The Meta Quest 2 VR HMD, along with the corresponding two wireless controllers, was chosen. It provides six degrees of freedom, allowing the user to move freely and easily. Head and hand movements were translated one-to-one in the virtual environment. Except for the embodiment scenario, the user could see virtual hands that represented their hand position and gestures made with the controllers. They did not have a full-body representation of the user (except in the \emph{positive} condition of the embodiment scenario). Since we maintained consistency between negative and positive embodiment and solely focused only on score differences, we believe that this limitation did not affect our results. The test scenarios were created using the Unity 3D game engine.\\

\begin{figure}[tb]
        \centering
            \includegraphics[width=\columnwidth]{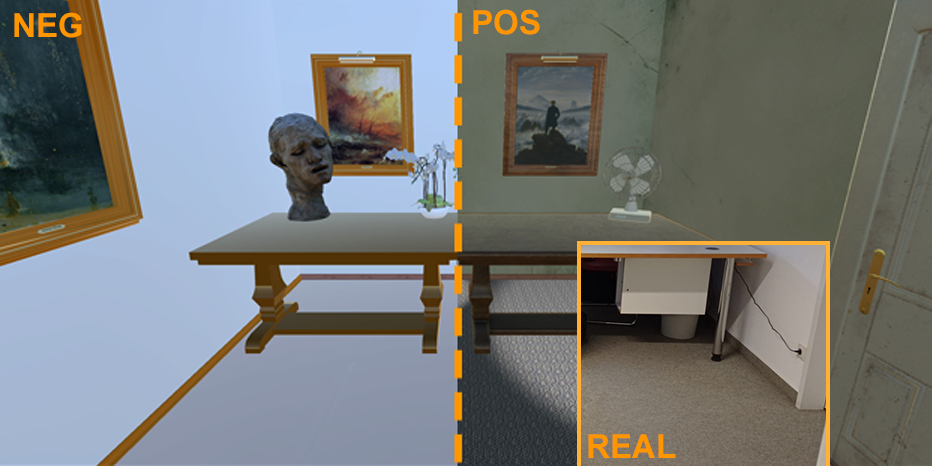} 
            \caption{The virtual environments of the place illusion scenario. On the left is the low-dimension intensity condition, and on the right is the high-dimension intensity condition. On the bottom right is a section of the real-world room from the same perspective.}
            \Description{Screenshots of the place illusion scenario. The scenario consisted of a virtual room. The observer is looking at a virtual table. Three images are hanging on the wall. A statue, a fan, and a flower are positioned on the table. There is a door on the right. The left screenshot shows the room with basic textures and without light rendering and shadow casting. The screenshot is labeled with "negative" at the top left. The right screenshot shows the room with detailed textures, light rendering, and shadow casting. It is labeled with "positive" on the top left. A photo on the bottom right of the image shows the actual study room, which has the same arrangement and dimensions as the virtual scenarios. It is labeled with "real" on the bottom left. A dotted line visually marks the boundary between the screenshots in the middle of the image. The right image shows the doors and the cabinet of the study room, which are again highlighted with orange labels each.}
            \label{fig:place_illusion_scenario}
        \end{figure}

\paragraph{Place illusion dimension}
Slater defines the dimension of \emph{place illusion} as "[\dots] the human response to a given level of immersion [\dots] bounded by the set of SCs possible at that level of immersion"~\cite{slater_place_2009}. In this context, sensorimotor contingencies (SCs) refer to learned actions that influence an individual's perception. Thus, we decided to modify the levels of immersion to achieve the desired changes in PI. In accordance with Slaters' concept of immersion ("[\dots] immersion is bound to a particular set of valid actions that support perception and effectual action within a particular virtual reality [\dots]"~\cite{slater_place_2009}), we designed the changes between the two conditions in terms of sensory breath and sensory depth, referring to the number and resolution of sensory channels for the user. For this purpose, we created a 3D model of the study room in its true-to-scale size. Real-world objects, such as the cabinet and the tables, were placed in identical positions in the VE to enable the participants to touch them (\autoref{fig:place_illusion_scenario}). The fan also existed in the real world and was running during the \emph{positive} condition to create a haptic sensation of wind. Instead of the flower, we placed a magnolia-scented candle on the table. It was uncovered during the \emph{positive} condition. All items were placed on the back of the table, out of the users' reach, to prevent them from trying to touch the objects. Otherwise, the haptic mismatch with the visual appearance of the objects may have irritated them. In addition, four paintings were placed on the wall to provide a visual task for the users. Three spatial audio sources were positioned in the VE to provide auditory stimulation that matched the visual experience.\\

\noindent To achieve changes in presence, we replaced all high-definition photo-textures with simple one-colored textures in the \emph{negative} condition. In addition, the directional spotlights positioned on the ceiling of the virtual room, shadow casting, and the fan were turned off in the \emph{negative} condition. The candle was covered with a glass. All audio sources were turned off. In both conditions, the users had two minutes to explore the room closely. The design challenge was to avoid any co-influences on other presence factor clusters. Therefore, no items were highlighted, and sounds were kept at a constant volume, which is usual for the type of sound, so that it would not affect unintended changes in involvement. No social perception was possible as no social partners were present. All objects that could be suggested to be interactive were placed out of the participants' range. Hence, we expected plausibility not to be affected by the task design. However, we cannot preclude effects of the additional sensory feedback from the \emph{positive} condition on plausibility. The users' virtual representation was equal for both conditions. Therefore, we did not expect user embodiment to be affected.\\

\begin{figure}[tb]
        \centering
            \includegraphics[width=\columnwidth]{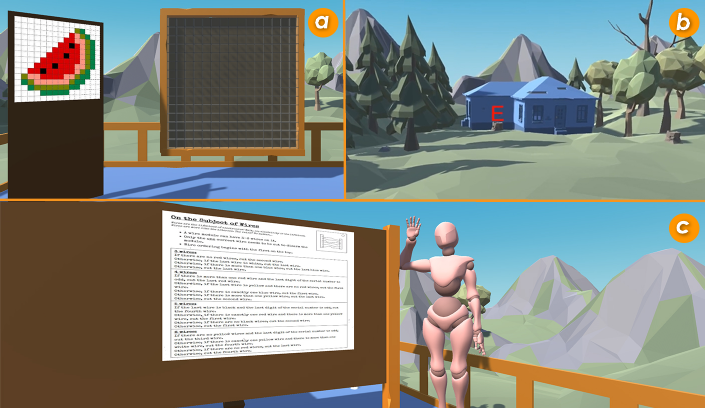} 
            \caption{The plausibility scenario (a), the involvement scenario (b), and the social presence scenario (c).}
            \Description{Screenshots of the plausibility, involvement and social presence scenario. The figure is composed of three distinct components, namely screenshots of the plausibility scenario (a) in the upper left, the involvement scenario (b) in the upper right, and the social presence illusion scenario (c) in the lower portion. The screenshot depicting the plausibility scenario showcases a low poly environment, with the observer situated on a fenced platform. In the immediate foreground, a pixelated depiction of a watermelon is positioned on the left, while a tiled wooden frame is situated directly ahead. This frame consists of approximately 16x16 fields, separated by a grid structure. The upper right image presents a low poly nature scenario, featuring a house positioned directly ahead of the observer. A stump is situated in front of the house, accompanied by a red letter that hovers above it. Trees are positioned on the left and right of the house, with mountains in the background. The screenshot located at the bottom shows a low poly environment, also with similar mountains in the background. The observer's vantage point is situated on a fenced platform, with a board positioned before him. A poster with text is affixed to the board, and an avatar, resembling a mannequin, is positioned adjacent to the board, gesturing towards the observer.}
            \label{fig:scenarios}
        \end{figure}

\paragraph{Involvement dimension}
We based the design of the involvement scenario on the definition of involvement by Witmer et al.~\cite{witmer_measuring_1994}, who defined it as "[\dots] focusing one's energy and attention on a coherent set of stimuli or meaningfully related activities and events". Thus, we focused the design on influencing the extent to which users paid attention to the aspects of the VE. We asked users to observe the VE and try to remember symbols that appeared for a short time (${\qty{2}{\second}}$). The VE was a landscape, as shown in~\autoref{fig:scenarios}. The users were positioned in a fenced area within the landscape so that they automatically stayed in the play area. In the \emph{negative} condition, exactly one letter appeared once in the virtual environment. In the \emph{positive} condition, five letters appeared one after the other, forming a word. Before the condition started, users were told how many symbols would appear and that the task was to remember them and enter them in an interface that appeared after 100 seconds. In the \emph{negative} condition, the symbol appeared after 10 seconds. Since the only task was to spot this single symbol, the users should focus less on the virtual environment once they recognize it. In the \emph{positive} condition, a symbol appeared after every 15 seconds. Consequently, we expected the users' involvement to be higher in the \emph{positive} condition, as they had to actively scan the VE until the end of the condition in order not to miss a symbol. 
The only differences between the conditions were the number of displayed symbols and the timing of the symbols' appearance. Hence, we did not expect any effects of the changes on place illusion, plausibility, embodiment, or social presence.\\

\paragraph{Embodiment dimension}
To test the effect of embodiment on the perceived presence of users, we used the approach of Zhang et al.~\cite{zhang_influence_2020}. For our investigations, we compared a full-body embodiment condition with a condition in which only the controllers represented the user's hands. In the \emph{positive} condition, interactions with the controllers were translated into gestures of the virtual hands, while the controllers themselves were not visualized. To mimic a real jigsaw puzzle, users could grab pieces, rotate them, and release them at the desired location. If the rotation and shape of the piece matched, the piece would fit into the corresponding slot.\\
The difficulty of the task remained the same because the interaction was unaffected. In both conditions, there were no changes in environmental detail, acoustic feedback, or task difficulty. Thus, changes in presence should be due to differences in embodiment. According to the results of Zhang et al., users should perceive more presence in the \emph{positive} condition~\cite{zhang_influence_2020}.\\

\paragraph{Social presence dimension}
We based our definition of the \emph{social dimension} of presence on Lombard's categorization regarding social realism and social presence~\cite{lombard_at_1997}. This includes parameters such as the feeling of communicating with an entity as a social partner. Therefore, we designed a basic avatar that resembled a human in stature but was kept minimalistic in detail. To manipulate social realism, we alternated between a computer-generated voice and a real voice and modified the interlocutor's conformity to social norms. To achieve interaction between user and avatar, we implemented a game similar to "Keep talking, and nobody explodes"\footnote{https://bombmanual.github.io/english/} (\autoref{fig:scenarios}). Users had three minutes to solve the task. At the beginning of each condition, the avatar stood to the side of users, greeted them, and then walked behind the board. After the riddle was solved, the avatar returned to the front of the board and said goodbye to the users. For the \emph{positive} condition, the supervisor himself replied to the users' questions. His voice was transmitted to the headphones to support the illusion of co-presence in the virtual world. The avatar first greeted the users, asked for their names, and responded accordingly. In the \emph{negative} condition, the avatars had a computer-generated voice that read texts entered by the supervisor or predefined responses that were unlikely to match the users' questions. Responses partially interrupted the users or were presented with delay. Environmental detail was not altered and was reduced to the minimum necessary for task completion. In both scenarios, the interaction was the same with respect to the task, and only the details of the social interaction changed between conditions. Since the task difficulty is the same in both conditions, we argue that there were also no differences in involvement between the conditions. The embodiment of the users was the same in both conditions.\\

\paragraph{Plausibility dimension}
To realize changes in plausibility, we designed a scenario where users were asked to recreate a pixel art painting. An empty canvas was placed next to the painting (\autoref{fig:scenarios}a). The users had to choose the right-colored cube by pressing left or right on their controller and aiming at the desired spot on the canvas. The task was the same in both scenarios. In the \emph{positive} condition, the interaction worked as explained by the experimenter. The color picker would sequentially go through the list of available colors and repeat when at the end of the list. In the \emph{negative} condition, the placement of the cubes was randomized. In addition, colors were randomly picked by the color picker, which could result in colors being picked multiple times. No social interaction was included, which could have influenced social interaction factors. The task was equivalent in both conditions to avoid possible differences in involvement. The interaction also remained the same in both conditions, with the expectation that identical actions could lead to different outcomes in the \emph{negative} condition. The changes should not have had any impact on the place illusion dimension, as the VE was the same in both conditions. However, we considered it important to provide interaction feedback through a designated sound when a cube was placed in a random location. Otherwise, users might have believed that the misplacement was due to an error. However, since correct placement was also supported by auditory feedback, we do not expect this design choice to affect the place illusion dimension.\\

\subsection{Procedure}
Overall, the study design resulted in three independent variables: (1) the \emph{presence dimension}, including place illusion, involvement, plausibility, social presence, and embodiment, (2) the \emph{dimension intensity}, which could be either low or high, and (3) the \emph{presence questionnaire} including the SUS, PQ, IPQ, and SIP. We used a within-study design and counterbalanced the order of scenarios and conditions per scenario across users according to the Latin square design. Thus, potential carry-over effects between the scenarios should have been counterbalanced.
None of the questionnaires provide a prior explanation of the concepts included, such as presence. Subsequently, we chose not to bias participants' judgments by defining presence or any related concepts to promote authentic usage. Due to the assumed positive effects on the reliability and consistency of presence assessments~\cite{schwind_using_2019,graf_incons_2020}, participants were asked to answer all questionnaires without leaving the VR. To avoid the presence ratings being influenced by the experience of the questionnaire assessment itself, we chose to use a neutral lobby scenario for it. 
The SUS and IPQ were rated on a 7-point Likert scale, the PQ was scored on a 5-point Likert scale, and the SIP was scored on a 10-point Likert scale using the response options provided by the questionnaires. To facilitate a comparison of the different scores, we converted the scores to an equal range between 1 and 7. 
All questions are summarized and roughly categorized in the appendix in~\autoref{tab:questionnaire_sus}-~\autoref{tab:questionnaire_bouchard}. At the end of the study, we also asked a few demographic questions.\\ 

\subsection{Participants}
Based on a \textit{a priori} power analysis (G*Power v3.1.9.7, repeated measures within-factors ANOVA, $\alpha=0.05,$~$power = 0.9,~r = 0.25$), the study was conducted with $N=50$ participants (26 male, 24 female). Participants were recruited from university students and staff members and had a mean age of $25.39$ years ranging from $18$ to $47$ years. $78.0\,\%$ had prior experience with VR of which $36.0\,\%$ only used it occasionally and $34.0\,\%$ used it for less than two hours per week.\\

\section{Results}
To ensure that our results are meaningful, we compared equivalent questions from the different questionnaires. If the participants answered conscientiously, the answers should be the same. We compared the PQ7\_consistency and IPQ12\_consistency (difference $M = 0.12$). The dataset did not show any inconsistencies.\\

    \subsection{Sensitivity of Presence Questionnaires}
\begin{figure*}[h]
                \centering
                    \includegraphics[width=\textwidth]{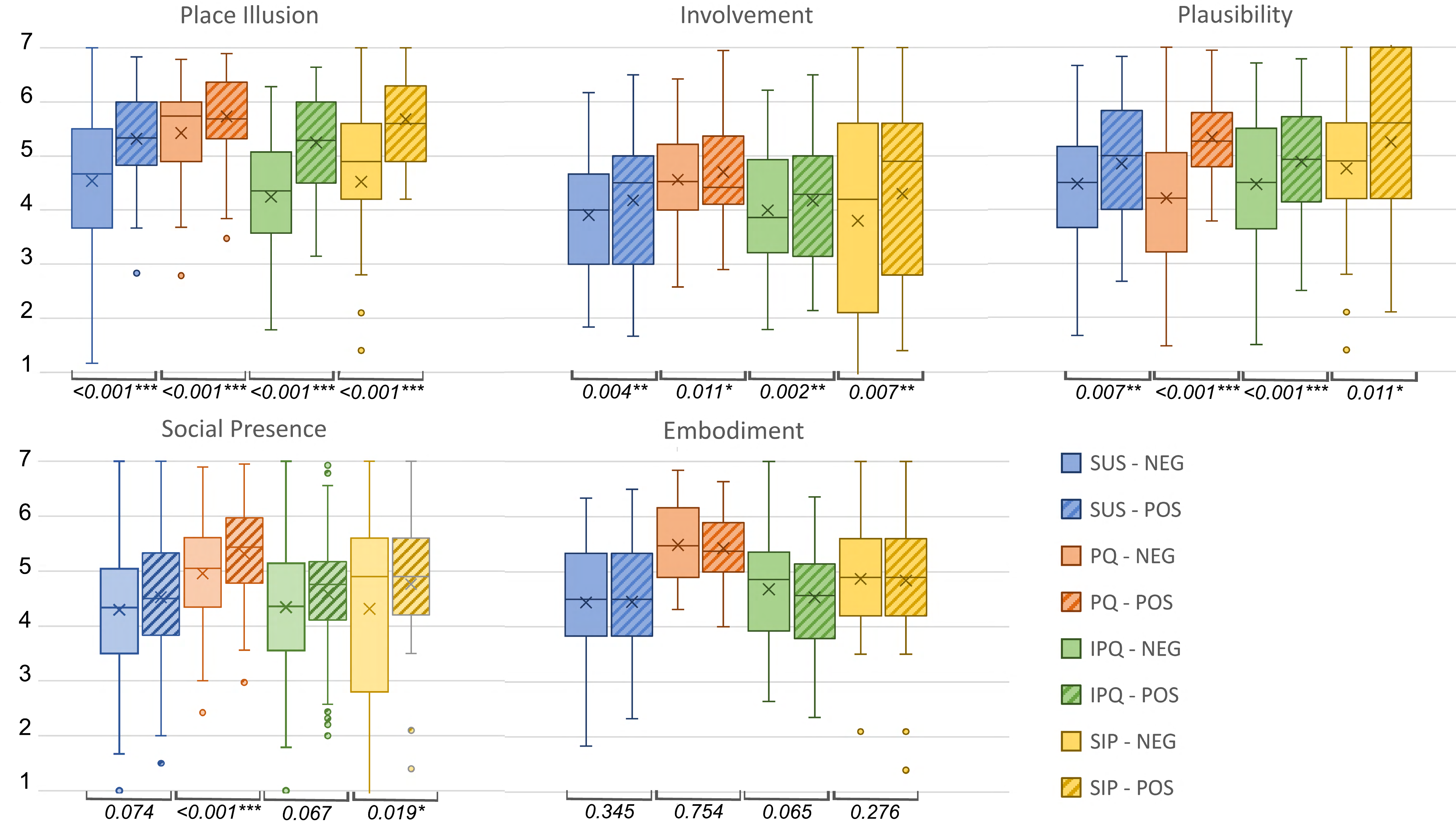}
                    \caption{Distribution of the measured scores from the SUS, PQ, IPQ, and SIP questionnaires. Each graph shows the scores from all four questionnaires for a given dimension (blue = SUS, red = PQ, green = IPQ, yellow = SIP) paired as negative (solid pattern) and positive (stripe pattern) conditions. The cross indicates the mean, and the line indicates the median.}
                    \Description{Boxplots of the scores of SUS, PQ, IPQ and SIP for negative and positive condition split by scenario. Mean, median and standard deviation of the scores per condition and dimension are also listed in Table 7 and Table 12. The place illusion plot demonstrates the highest scores (mean between 4.5 and 5.7) with comparatively concentrated distributions (interquartile range [IQR] around 1), with the exception of the negative scores of SUS, which exhibit an IQR of 1.8. With regard to the involvement dimension, scores are more widely distributed, and the highest and most concentrated distributions are exhibited by the PQ (negative M = 4.44; positive M = 4.66). The most widespread score distributions of the involvement dimensions can be observed for the SIP scores (negative M = 1.73, IQR = 3.46; positive M = 1.80, IQR = 1.5). The distribution patterns for the plausibility and social presence dimensions exhibited a more even distribution, with means ranging from 4.16 to 4.90 for the plausibility dimension and from 4.29 to 5.32 for the social presence dimension. The embodiment dimension demonstrated the most uniform distribution, with the highest scores observed for the PQ and IQRs ranging from 1 to 1 for all conditions and questionnaires. Statistical tests revealed significant differences between negative and positive scores for all questionnaires in the place illusion (p < 0.001), involvement (p < 0.011), and plausibility dimension (p < 0.011). However, only the PQ and SIP scores exhibited significant differences for changes in the social presence dimension, and none of the questionnaires exhibited significant differences for the embodiment changes (p > 0.065).}
                \label{fig:scores_perScenario}
            \end{figure*}

\begin{figure}[h]
            \centering
                \includegraphics[width=\columnwidth]{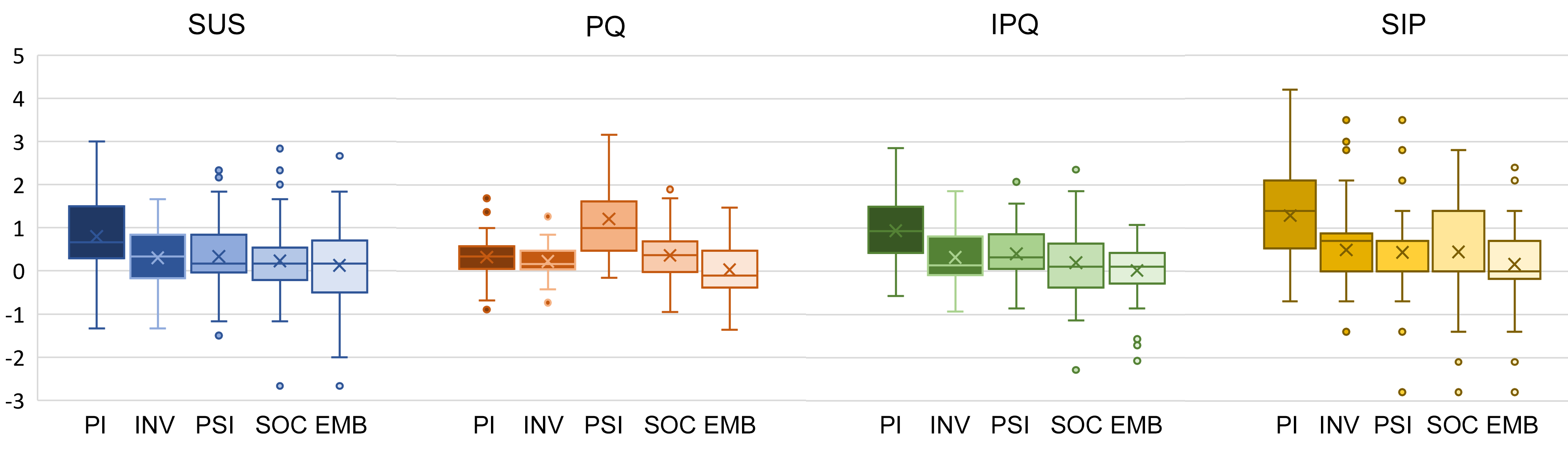}
                \caption{Distribution of the score differences (score of positive condition - score of negative condition per user) of SUS, PQ, IPQ, and SIP clustered by dimension (left = place illusion, second left = involvement, middle = plausibility, second right = social presence, right = embodiment). The line indicates the median value, and the cross indicates the mean.}
                \Description{Boxplots of score differences per questionnaire and scenario. Four boxplot groups with five individual boxplots, one for each boxplot. The y-axis indicates the score difference, ranging from -3 to 5, with each scenario-questionnaire combination represented by a boxplot grouped by questionnaires. The score differences are predominantly slightly positive on average, with means ranging from -0.15 to 0.73. However, certain questionnaires demonstrate more pronounced score differences, such as the SUS for the place illusion scenario, with a mean of 0.80, the P Q for the plausibility scenario, with a mean of 1.21; the IPQ for the place illusion scenario, with a mean of 0.94; and the SIP again for the place illusion scenario, with a mean of 1.29.The distributions of the score differences for SUS and SIP are more widespread than the distributions for PQ and IPQ. Outliers occur for all questionnaires, with most exhibited by the SIP.}
                \label{fig:score differences per questionnaire and scenario}
        \end{figure}

    To address our first research question, "How sensitive are they to changes of specific dimensions of presence?" we compared the scores of the positive and negative conditions for each scenario and each questionnaire. The results are shown in~\autoref{fig:scores_perScenario}. The average of the presence scores for each scenario and questionnaire evidenced an upward trend from the negative to the positive condition (differences between $M=0.02$ and $M=1.29$ on average). As shown in ~\autoref{fig:score differences per questionnaire and scenario}, all questionnaires result in more substantial score differences between the conditions in the \emph{place illusion} dimension. The SUS, IPQ and SIP have their largest average difference in this scenario (SUS: $M=0.80$, IPQ: $M=0.94$, SIP: $M=1.29$). The greatest difference of the PQ was found in the \emph{plausibility} scenario. For all questionnaires, the \emph{embodiment} scenario performed the lowest with an average score difference ranging from $M=0.02$ to $0.016$. Except for the EMB and PI scenarios, the SUS, IPQ, and SIP had comparably similar score differences to the other three scenarios (SUS: $M=0.29\pm 0.07$, PQ: $M=0.31\pm0.11$, SIP: $M=0.46\pm0.03$). The PQ showed comparable results across the PI, INV, and SOC scenarios, with a mean of $0.3\pm0.08$.

        \noindent We performed \emph{related-samples Wilcoxon signed-rank} tests to compare the conditions within each scenario for each questionnaire. For further insight, we determined the respective effect sizes (small effect: $0.1-\leq 0.3$, medium: $0.3-\leq 0.5$, large: $\geq 0.5$) and 95\,\% confidence intervals of the differences between positive and negative scores using the \emph{Hodges-Lehman estimator} for non-parametric related samples~\cite{newson2002, wasserstein2016}. \\
        The results are summarized in~\autoref{fig:scores_perScenario}. For further information on the results, refer to~\autoref{tab:scoredifferencewilcoxon} in the appendix. The results of the dimension-specific evaluation of the PQ and IPQ are presented in table \ref{tab:subscales_diff}.\\

    \subsection{Approval of Sensitivity}
\begin{figure*}[h]
        \centering
            \includegraphics[width=\textwidth]{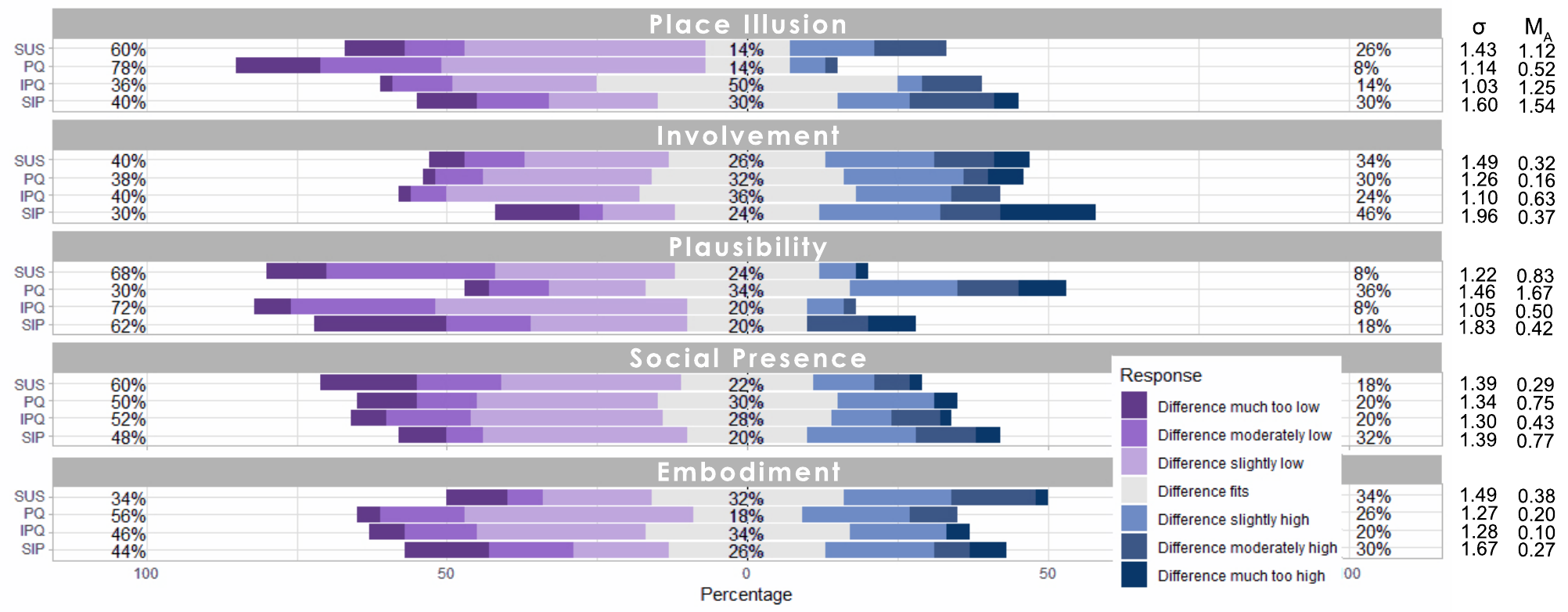}
            \caption{Participants' agreement with the score differences of the different questionnaires per scenario. 
            Participants could rate the score differences on a scale from -3 (purple) = "Difference is much too low" to 3 (blue) = "Difference is much too high", where 0 (gray) = "Agree with the score difference".}\label{fig:presence_approval_scoredifferences}
            \Description{Horizontal stacked bar chart of user approval to questionnaire scores grouped by dimension. Each bar represents a certain questionnaire, and every color within the bar represents the frequency of a certain approval score. Percentages on the left and right, as well as in the middle of the bar, indicate the proportion of approval to score differences being perceived as too high (right) or too low (left), respectively.}
        \end{figure*}

    \noindent Participants rated the score differences for each questionnaire on a scale from $-3=$ "Difference is much too low" to $3=$ "Difference is much too high," with $0=$ "Agree to the score difference." The results are illustrated in~\autoref{fig:presence_approval_scoredifferences}.\\
    The results showed that the IPQ scores aligned most closely with users' expectations ($|M|=0.964$, $\sigma=1.21$), with $33.6\%$ of participants agreeing with the score differences. The PQ achieved the second-highest agreement rate ($25.6\%$). Across scenarios, participants generally perceived score differences as too low, except for the involvement and plausibility scenarios, where some questionnaires had higher-than-expected scores (e.g., INV: SIP: $M=0.26$, $\sigma=1.87$).

    \noindent The SUS and PQ had the lowest reflection of changes users agreed with. The SUS aligned most closely with user experience in the embodiment ($M=-0.08$, $\sigma=1.49$) and involvement scenarios ($M=-0.06$, $\sigma=1.52$). For the PQ, the highest agreement was in the involvement scenario ($M=-0.04$, $\sigma=1.3$) and the lowest in the place illusion scenario ($M=-1.16$, $\sigma=1.14$). The IPQ had the highest agreement in the involvement scenario ($M=-0.18$, $\sigma=1.09$) and the lowest for plausibility ($M=-0.98$, $\sigma=1.05$).

    \noindent Variance was highest for SIP scores, particularly in the social presence scenario ($\sigma=1.73$), while the IPQ showed the lowest variance across all scenarios ($\sigma=0.89$).  The results are summarized in \autoref{tab:scores_approval}, \autoref{tab:approval_overview} and \autoref{tab:approval_overview_conditions}. Non-parametric Friedman tests indicated significant differences in the user agreement for the place illusion ($\chi^2(2)=23.74$, $p<.001$) and plausibility ($\chi^2(2)=30.92$, $p<.001$) scenarios. Post-hoc analyses revealed significant differences in agreement with PQ score differences compared to other questionnaires ($p\leq.018$ for place illusion, $p<.001$ for plausibility).
    
\section{Discussion}
    
    The results showed that our approach to investigate dimension-specific sensitivity works. All questionnaires showed an upward trend in scores from negative to positive conditions for all presence dimensions except for the embodiment scenario. In compliance with previous work, we found a significant bias of presence assessments introduced by design. While previous work showed a bias caused by the questionnaire choice~\cite{Hein2018TheUO}, the assessments' setting~\cite{graf_incons_2020} and individual interpretation of questionnaire items~\cite{usoh_using_2000}, we focus our analysis on the effects caused by changes in specific presence dimensions. The reason is that presence questionnaires by design are not considered to be suitable for cross-scenario comparisons but rather for direct comparisons of VR experiences. Accordingly, Schwind et al. assume that the questionnaires do not directly measure the construct of presence but the contrast of VR experiences \cite{schwind_using_2019}. Hence, a VR experience could also receive lower presence ratings than another, although the user was comparably present. Simply caused by the fact that an aspect of the experience was perceived as less qualitative than in the other experience, regardless of whether this discrepancy influenced presence or not. Compared to another VR experience, the score might turn out differently. Hence, we not only evaluate the questionnaires' sensitivity but also users' approval of the resulting scores. In fact, not only did the dimension-specific sensitivity analysis provide meaningful insights regarding differing sensitivity to changes in the respective dimension, but the assessment of user approval proved that greater sensitivity does not necessarily mean more authentic presence assessments. 

    To address our research questions RQ1 ("How sensitive are the questionnaires to changes of specific dimensions of presence?"), RQ2 ("Do participants agree to the differences in scores affected by the changes of a specific dimension of presence?"), and RQ3 ("Can questionnaire items be clustered according to their sensitivity to changes of a specific dimension of presence?"), we conducted a comprehensive analysis of the quantitative results of our study. This included an overall analysis (RQ1), an examination of participants' approval of scores (RQ2), and an analysis of scores per scenario as well as per question (RQ3). Statistically significant differences were observed between the scores of the positive and negative conditions for certain scenarios related to a particular questionnaire. However, although we designed the embodiment scenario based on previous research, the variations between the respective conditions were not significant enough to achieve measurable differences in presence. Participants agreed that the differences between the two conditions were negligible. Consequently, we decided to omit this dimension from further discussion.\\

    \subsection{Sensitivity of Presence Questionnaires} 
    Our results demonstrate the influence of a specific presence dimension on presence scores, with higher scores on average in the positive condition. However, there were discrepancies between the presence questionnaires and dimensions. The \emph{SIP} questionnaire showed the greatest sensitivity to change, with a mean difference of $0.57$ in scores. The \emph{SUS} had an average change of $0.36$, while the \emph{PQ} and \emph{IPQ} had changes of $M=0.38$ and $M=0.43$, respectively.
    The \emph{SIP} scores showed increased differences, possibly due to changes not moderated through averaging multiple questions, some of which may partially not have been triggered by the changes. However, the high standard deviation ($\sigma=30.9\,\%$ higher than average with $\sigma=1.25$) and the lower user agreement ($11.4\,\%$ below average) indicate that users have difficulty quantifying perceived presence.\\
    \noindent The \emph{SUS} showed the highest sensitivity to the changes made in the place illusion scenario, probably because it primarily considers the perception of presence in general as well as environmental aspects. The sensitivity to changes in the other presence dimensions was lower and comparably uniform across the scenarios. Apart from the SIP, the SUS exhibited the second-highest sensitivity to the variations in the involvement scenario. \\
    \noindent The \emph{IPQ} demonstrated a comparable sensitivity profile, with the greatest sensitivity in the place illusion scenario, which was slightly higher than that of the SUS and lower sensitivity in the other scenarios. The IPQ was slightly more sensitive to changes in the PSI scenario but less sensitive in the involvement scenario. 
    Both questionnaires appear to be suitable for environmental aspects-focused scenarios but provide moderate sensitivity to changes in the other dimensions. The IPQ is preferable for cases with respect to plausibility, while the SUS is better able to discern differences in users' involvement.\\
    \noindent The \emph{PQ} was less responsive to variations in involvement and place illusion. With an average score difference of $M=0.32$, he was the least sensitive questionnaire in the place illusion scenario. The PQ showed the highest sensitivity towards alterations in the plausibility scenario ($M=1.21$). According to Slaters' definition of plausibility~\cite{slater_place_2009}, the alterations in plausibility are likely to have had an impact on naturalness evaluated by multiple items. They may also have contributed to the PQ being the most responsive questionnaire to the modifications in the social presence scenario since the interaction with the avatar and the reception of real voice replies should enhance naturalness. None of the questionnaires demonstrated significant sensitivity in the social presence scenario, probably because none considered respective aspects. Another questionnaire, such as the one created by Nowak and Biocca~\cite{nowak2003}, which considers social aspects, may have performed better in this situation. The analysis of the individual dimensions of PQ and IPQ showed that the changes made to the involvement scenario did not impact the relevant items in the PQ as expected. We speculate that the minor effects can be attributed to our focus on involvement with visual aspects of the VE as opposed to the various involvement types considered by the PQ's items on involvement. Thus, the individual use case influences the sensitivity of certain dimensions of the PQ. Probably due to the involvement items mainly focusing on how VE design affects the user's focus rather than on the design itself, as the PQ items do, the IPQ showed greater sensitivity to changes in the involvement scenario. We anticipated greater sensitivity to the changes in the social presence and embodiment scenario since the positive conditions should be more realistic. We assume the changes to be too specific, as the items primarily concern the VE's realism in general, and the VE's of the respective scenarios were still comparatively unrealistic. In conclusion, the sensitivity of the PQ's and IPQ's items strongly depends on the associated use case and the use of another questionnaire that considers this dimension may already be more or less sensitive.\\

    \subsection{Approval of Sensitivity} 
    We presented the scores of both conditions of each scenario to the participants, who were then asked to indicate their level of approval regarding the score differences based on their experience. Overall, the participants would have expected a greater score difference in almost all cases. The differences observed in \emph{IPQ} scores were most consistent with their expectations, demonstrating a $12\,\%$ higher level of approval than the other questionnaire approvals. A closer examination of approval ratings per scenario indicates that participants perceived greater differences between the conditions of the place illusion scenario than reflected by the score differences. The IPQ and SIP proved most sensitive in this scenario, although participants agreed more with the differences in the IPQ scores. We attribute this effect to the high variance in \emph{SIP} scores (SIP: $\sigma=1.6$, IPQ: $\sigma=1.09$). Based on the higher frequency of "much too low" and "much too high" ratings, $54.54\,\%$ more frequent than in the other questionnaires, the single item concept may lead to under- or overestimating perceived presence. On average, the SIP score difference received the lowest approval ratings on average ($|M|=1.384$), and both the score differences ($\sigma=1.25$) and approval ratings ($\sigma=1.73$) had a high standard deviation.
    The approval to the differences in scores for the \emph{PQ} varied significantly ($p<.001$). The lowest agreement was reported for the PI scenario ($M=-1.16$), where score differences were also the lowest (PI: $M=0.84$, PI\_PQ: $M=0.32$). In contrast, users mostly agreed on the score difference of the PQ in the plausibility scenario ($M=0.14$). 
    Alongside the SIP, the SUS was the second most sensitive questionnaire in the INV scenario and received the most user approval. The PI scenario showed the greatest sensitivity. However, users still scored the differences too low in most cases. Regarding the social presence scenario, none of the questionnaires satisfied the users' expectations. On average, all score differences were considered too low ($M=-0.404$). This aligns with our expectations, as none of the questionnaires includes items related to social aspects.\\

   \noindent We did not determine the users' approval of the score differences in the individual presence dimensions of PQ and IPQ. Nevertheless, we conducted a brief comparison of their score differences with the differences of the overall score differences to get insights as to whether participants would have potentially concurred more with the score discrepancies of the items linked to the respective presence dimension. Participants could have indeed agreed more on score differences of specific dimensions as, in some cases, these showed a greater difference between the two conditions. The changes in the plausibility scenario affected items on involvement and naturalness. The changes of place illusion had mixed effects. As anticipated, the items on spatial presence, involvement, and realness of the IPQ were affected. However, we would have expected greater effects on the resolution items of the PQ. Moreover, the questions about the involvement of both PQ and IPQ were less impacted by the variations in the involvement rating than anticipated. Therefore, a follow-up study that investigates the sensitivity of questionnaire items or subscales could offer valuable insights. Introducing weighting on items pertinent to a particular usage scenario may potentially enhance the alignment of score variations with user experience.\\
    
    \subsection{Sensitivity of Single Items}
    We also analyzed items that were less affected or less affected by change than anticipated. 
    As illustrated in table \ref{tab:subscales_diff}, items on involvement showed little sensitivity in all scenarios. Most effects on involvement were observable for the place illusion scenario. We assume the limited sensitivity was due to participants' difficulties in interpreting the term involvement, as we were repeatedly asked about its meaning. This assumption is supported by the fact that the \hyperlink{IPQ7}{\textit{IPQ7}} and \hyperlink{IPQ8}{\textit{IPQ8}}, which ask the user whether they are paying attention to the real world, were more affected by the changes.

    Another question that received low scores regardless of the scenario was \hyperlink{IPQ14}{\textit{question 14 of the IPQ}}. This question asks the user whether the virtual world is more realistic than the real world. As expected, this question was rated very low for both conditions in each scenario (positive: $M=2.32$, negative: $M=2.12$, difference: $M=0.192$). We question the relevance of this question, as previous research~\cite{cheng2005immersionBarriers,jicol_realism_2023} suggests that
    presence does not necessarily depend on high realism. Slater and Lee both argued that instead of high realism, it is only necessary for the user to be able to construct a mental model that accounts for what they perceive in the virtual world~\cite{slater_place_2009, lee_presence_2004}. Baños proposed that presence and reality judgments are two separate concepts that may or may not be positively correlated with each other but do not have to~\cite{banos2000}. The relevance of realism for presence depends on the individual context of use. Furthermore, we believe that some users may not even know how to define a world that is more realistic than what we know as reality.\\

    \section{Limitations}
   Since the PQ and IPQ include items on involvement, we would have expected greater effects of our changes in the INV scenario. We designed our scenarios based on Witmers' and Singers' definition of involvement~\cite{witmer_measuring_1994}. It would be interesting to examine how different types of involvement affect larger score differences. Our embodiment scenario design was based on previous work. However, the differences between the conditions were too small to achieve perceivable differences. We consider the short task duration or the ease of the task to have potentially caused minor effects. A follow-up study on the sensitivity of the questionnaires to changes in social realism and user embodiment would be interesting to determine if the small score differences were indeed caused by the limited sensitivity of the questionnaires or by the design of our conditions.

    \section{Implications}
    \begin{table}[ht]
    \centering
    \caption{Mean score differences (positive minus negative score per user) including significance level and mean approval ratings of questionnaires per presence dimension (dim.) and questionnaire (que.) (PI = Place Illusion, INV = Involvement, PSI = Plausibility, SOC = Social presence and EMB = Embodiment).}
    \label{tab:sensitivity_approval}
    \Description{Overview of the results of sensitivity and approval analysis. The results are ordered by dimension and questionnaire, stating average score difference with indicators for the significance level of the respective statistical test, standard deviation of the score differences and user approval on average as well as its standard deviation.}
    \begin{tabular}{l|c|c|c}
    \hline
    \textbf{Dim.}      & \textbf{Que.} & \textbf{Score Difference} & \textbf{Approval} \\ \hline
    \textbf{PI} 
      & SUS  & $0.77$**$~(\sigma= 0.90)$ & $-0.52 ~(\sigma=1.19)$\\ 
      & PQ   & $0.31$**$~(\sigma=0.50)$    & $-1.16  ~(\sigma=0.46)$\\ 
      & IPQ  & $0.88$**$~(\sigma=0.88)$    & $-0.26 ~(\sigma=1.37)$\\ 
      & SIP  & $1.15$**$~(\sigma=1.15)$    & $-0.20~(\sigma=1.24)$\\ \hline
    \textbf{INV}
      & SUS  & $0.32$*$~(\sigma=0.74)$ & $-0.06 ~(\sigma=0.38)$\\ 
      & PQ   & $0.22~(\sigma=0.63)$ & $-0.04~(\sigma=0.18)$\\ 
      & IPQ  & $0.29$*$~(\sigma=0.32)$ & $-0.18  ~(\sigma=0.71)$\\ 
      & SIP  & $0.55$*$~(\sigma=0.52)$ & $0.26~(\sigma=0.30)$\\ \hline
    \textbf{PSI} 
      & SUS  & $0.33~(\sigma=0.80)$ & $-1.04~(\sigma=0.50)$ \\ 
      & PQ   & $1.17$**$~(\sigma=0.88)$ & $0.14~(\sigma=0.27)$\\ 
      & IPQ  & $0.40$*$~(\sigma=0.40)$ & $-0.98~(\sigma= 0.50)$\\ 
      & SIP  & $0.47~(\sigma=0.47)$ & $-0.76~(\sigma= 0.73)$\\ \hline
    \textbf{SOC}
      & SUS  & $0.23~(\sigma=0.93)$ & $-0.78~(\sigma=0.71)$ \\ 
      & PQ   & $0.38$**$~(\sigma=0.62)$ & $-0.52~(\sigma=1.73)$ \\ 
      & IPQ  & $0.22~(\sigma=0.04)$ & $-0.46~(\sigma=0.45)$ \\ 
      & SIP  & $0.45~(\sigma=0.24)$ & $-0.20~(\sigma=0.60)$ \\ \hline
    \textbf{EMB}
      & SUS  & $0.14~(\sigma=0.95)$ & $-0.08~(\sigma=0.31)$ \\ 
      & PQ   & $-0.01~(\sigma=0.62)$& $-0.44~(\sigma=1.27)$ \\ 
      & IPQ  & $0.05~(\sigma=0.02)$ & $-0.42~(\sigma=0.59)$ \\ 
      & SIP  & $0.09~(\sigma=0.08)$ & $-0.38~(\sigma=0.47)$ \\ \hline
    \end{tabular}
    \end{table}

    Our study was the first to quantify the yet only assumed bias of presence assessments by the choice of a questionnaire. Even though our scenarios were designed to maximize dimension-specific differences, the questionnaires' score differences were comparatively small, and users would have expected greater score differences in almost all cases (\autoref{tab:sensitivity_approval}). Furthermore, the dimension-specific sensitivity varied significantly among the questionnaires. As VR scenarios are rarely designed to match certain conceptual requirements of presence but concern a variety of presence dimensions, we assume that the bias is usually even larger as it is more likely that the chosen questionnaire does not perfectly cover the presence dimensions affected by the scenario. Arguably, requesting a more concept-focused scenario design for presence investigations is not realistic. To promote the validity of presence assessments, awareness of such choice-dependent biases should be raised on the side of presence researchers, and we strongly encourage questionnaire developers to provide more detailed guidelines on the applicability of questionnaires for certain presence dimensions. Based on the results of our sensitivity evaluation, we were able to provide various insights on the applicability of SUS, PQ, IPQ, and SIP:\\
    
    \textbf{Dimension-specific insights:}
    \begin{itemize}
        \item \textbf{Place Illusion:} Based on our results, we recommend using the IPQ or SIP. Both exhibited great sensitivity to the changes and received the most approval from the users. However, we recommend interpreting results cautiously as both also showed great uncertainty in terms of high standard deviation.
        \item \textbf{Involvement:} All questionnaires showed to be sensitive to changes in involvement, but the SUS and PQ seemed to reflect users' experience most accurately.
        \item \textbf{Plausibility:} The PQ showed exceptional sensitivity and was most approved by the users.  
        \item \textbf{Social Presence:} Users approved most of the SIP that was most sensitive to the dimension changes. However, sensitivity was still rated as too low. We recommend the choice of a questionnaire that explicitly considers social presence for investigations on that dimension.
        \item \textbf{Embodiment:} All questionnaires exhibited little sensitivity to the changes. However, according to user approval, the SUS could reflect the experience accurately.
    \end{itemize}

    \textbf{Questionnaire-specific insights:}
    \begin{itemize}
        \item \textbf{Slater-Usoh-Steed Questionnaire (SUS)~\cite{slater_virtual_2000,usoh_using_2000}: } The SUS showed to be a good choice when it comes to investigations on involvement and embodiment in terms of sensitivity and user approval. 
        \item \textbf{IGroup Presence Questionnaire (IPQ)~\cite{schubert_experience_2001,schubert_sense_2003}: } Based on our results, it seems that the IPQ is the best choice for VR experience that affect multiple presence dimensions as the questionnaire showed sensitivity to all of the dimensions. Especially if investigations consider place illusion, the IPQ might be a beneficial choice. 
        \item \textbf{Presence Questionnaire (PQ)~\cite{witmer_measuring_1994}: } The questionnaire was the least sensitive on average. We assume that this was caused by the interface quality subscale of the PQ as technical aspects were not changed between the conditions (e.g., \hyperlink{PQ17}{PQ17\_display quality}). Furthermore, we observed little sensitivity of the involvement subscale. We assume this to be caused by uncertainties due to the wording of the items. However, the PQ performed exceptionally well for changes in plausibility and was most approved by the users. This was likely caused by the PQ considering task-related aspects. Hence, we recommend the PQ for use cases focusing on task design.
        \item \textbf{Single-Item Presence Questionnaire (SIP)~\cite{bouchard_reliability_2004}: } The SIP exhibited the greatest sensitivity of all questionnaires. However, we observed uncertainty in the quantification of presence, which was illustrated by very mixed scores with equally mixed approval ratings. The SIP may be beneficial in measuring minor alterations in VR experiences. However, presence researchers should interpret differences cautiously, as they may not accurately represent users' actual experience.
    \end{itemize}

    \section{Conclusion}
    Presence research is often characterized by studies comparing different versions of a virtual experience to investigate the effects of alterations in design on presence. Hence, measurement tools are needed that are capable of accurately reflecting these effects on presence perception. Questionnaires are the most common way to assess presence. However, the choice of presence questionnaires faces some limitations: there is little research on (1) the sensitivity of questionnaires to changes in specific dimensions of presence and (2) the consistency of the resulting scores with users' actual experience. 
    Our work is the first to quantify the potential bias of presence assessments through the choice of questionnaires. We designed five virtual environments aimed at isolating changes in virtual reality that address a specific dimension of presence (place illusion, involvement, plausibility, social presence, and embodiment). Each scenario was implemented with two conditions (one to induce higher and one to induce lower presence). Analysis of the SUS, IPQ, PQ, and SIP scores from the low-dimension intensity and high-dimension intensity conditions revealed that the four questionnaires have different sensitivities to different dimensions of presence. The SUS and IPQ questionnaires were most sensitive to our place illusion scenario. The PQ showed the highest sensitivity to changes in plausibility. \\

    \noindent Our findings are consistent with our initial assumptions that not every questionnaire may be able to measure all aspects of presence with equal sensitivity. While all of them seemed to address aspects of place illusion and plausibility, the questionnaires showed low sensitivity when it came to aspects of user involvement, social presence, and user embodiment, with almost no sensitivity to the differences between the two conditions. Furthermore, we were able to point out that for most of the cases, the sensitivity of the scores was lower than expected by the participants. In particular, the discrepancies in the SUS scores indicated insufficient sensitivity to the changes. Bouchard's single-item presence scale showed comparably high sensitivity for all dimensions considered. Participants seemed to have difficulty in quantifying their perceived presence, resulting in high score variability and discrepancies with actual perception. As a result, the SIP questionnaire received the least agreement out of all four questionnaires despite being the most sensitive in all scenarios.
    Some questions in the IPQ appeared to cause difficulties in interpretation, resulting in lower scores, which may have led to lower approval ratings regardless of the altered presence dimension. Regarding the dimensions of the PQ and IPQ, some showed less sensitivity to changes in certain presence dimensions than expected. 
    Nevertheless, the results suggest that in some cases, participants could have approved more of the differences of the questionnaire items of that dimension. In general, our findings confirm our assumptions that the currently commonly used questionnaires are not as universally applicable as it seems to be assumed. \\
    
    \noindent Our approach of evaluating questionnaires' sensitivity to changes in specific dimensions of presence was able to point out a lack of sensitivity of the questionnaires in general but also for specific items that may be responsible for the overall lack of sensitivity. The results underline the need for awareness of the bias in presence assessments caused by the choice of a questionnaire on the side of presence researchers. Results should be interpreted cautiously as they might not reflect the users' actual experience. Second, as VR experiences are mostly not designed to match a certain concept of presence and often include aspects of various presence dimensions, we need more insights into a questionnaire's sensitivity to promote the validity of presence assessments. We encourage questionnaire developers to provide recommendations for the applicability concerning specific dimensions of presence, including insights on sensitivity and users' approval. This way, a more user-centered and case-sensitive presence assessment could be achieved, which would, in turn, increase the assessment's validity.\\

\begin{acks}
This research was funded by the Deutsche Forschungsgemeinschaft (DFG, German Research Foundation) through the project "Empirical Assessment of Presence and Immersion in Augmented and Virtual Realities" (project number: 425365472).
\end{acks}

\clearpage

\bibliographystyle{ACM-Reference-Format}

\bibliography{bibliography}

\clearpage
\onecolumn
\appendix

\section{Questionnaires}
    \subsection{Overview}
    \begin{table}[h]
        \centering
        \begin{tabular}{l c >{\centering\arraybackslash}p{0.7cm} p{70mm}}
            Authors & Year & Items & Factors/Indicators/Dimension \\\hline
             Banos et al. (\cite{banos_presence_2000}) & 1998 &  77 & Reality judgment, presence, emotional involvement, interaction, control, attention, realism, perceptual congruence, anticipation\\
             Barfield \& Hendrix (\cite{barfield_effect_1995}) & 1995 &  13 & Presence, navigation, real-world dissociation\\
             Bouchard (SIP~\cite{bouchard_reliability_2004}) & 2004 &  1 & None\\
             Dinh et al. (\cite{dinh_evaluating_1999}) & 1999 &  14 & Presence, involvement, realism, control, enjoyment\\
             Jennett (IEQ~\cite{jennett_measuring_2008}) & 2008 &  31 & Cognitive involvement, Real World Dissociation, Transportation, Challenge, Emotional involvement, Enjoyment.\\
             Kim \& Biocca (\cite{kim_telepresence_1997}) & 1997 &  8 & Arrival, departure\\
             Larsson et al. (SVUP~\cite{larsson_actor_2001}) & 2001 & 150 & Quality evaluations, attitudes, presence, realism, provided information, simulator sickness\\ 
             Lessiter et al. (ITQ-SOPI~\cite{lessiter_cross-media_2001}) & 2001 &  44 & Sense of Physical Space, Engagement, Naturalness\\
             Lombard et al. (TPI~\cite{lombard2009measuring}) & 2009 & 42 & Transportation, Immersion, Realism; Social Actor within a Medium Social Richness \\
             Lin et al. (E$^2$I~\cite{lin_effects_2002}) & 2002 &  14 & Sensory, distraction, realism, and control \\
             Nowak \& Biocca~\cite{nowak2003} & 2003 & 30 & Co-presence of self and others, telepresence, social presence\\
             Regenbrecht et al. (IPQ~\cite{schubert_experience_2001}) & 2001 &  14 & Spatial Presence, Involvement, Realness\\
             Slater, Usoh, Steed (SUS~\cite{slater_virtual_2000}) & 2000 &  6 & None \\
             Slater et al. (UCL~\cite{slater_depth_1994}) & 1994 & 3 & Subjective presence\\ 
             Takatalo et al. (EVEQ~\cite{takatalo_presence_2002}) & 2002 & 124 & Spatial presence, action, attention, real[ness], and arousal \\ 
             Vorderer et al. (MEC-SPQ~\cite{vorderer_mec_2004}) & 2004 &  L:72 M:54 S:36 & Attention allocation; Spatial situation model; Self-location; Possible actions; Cognitive involvement; Suspension of disbelief; Domain-specific interest; Visual/spatial imagery; absorption \\
             Witmer et al. (PQ~\cite{witmer1998measuring}) & 1998 & 32 & Involved/ Control, Natural; Interface quality\\\hline
        \end{tabular}
        \caption{Overview of presence questionnaires considered for this study with year of publication, number of items and dimensions included.}
        \Description{Overview of existing presence questionnaires considered for this work. The left column names the authors, title and reference of presence questionnaires. The second column shows the year of publication, the third the number of items, and the fourth column lists all factors/dimensions that are known to be considered by the questionnaire.}
        \label{tab:questionnaires_overview}
    \end{table}

    \clearpage

    \subsection{Items of the Questionnaires Used for Our Investigations}
    \begin{table}[h]
        \centering
        \begin{tabular}{| l p{8cm} l |}\hline
            \multicolumn{3}{|l|}{\textbf{Slater-Usoh-Steed Questionnaire (SUS, 2000)~\cite{slater_virtual_2000,usoh_using_2000}}}\\
            Question & Description                                                          & Tag \\ \hline
             1     & Rate the sense of being there.                                        & SUS1\_presence overall \label{SUS1}\\
             2     & Extent of VE being reality.                                           & SUS2\_became reality \label{SUS2}\\
             3     & Does user remember VE more as images or as environment?           & SUS3\_images \label{SUS3}\\
             4     & Was the sense of being in the VE or in real world stronger?  & SUS4\_dominant location \label{SUS4}\\
             5     & How detailed is the memory of the VE?                                 & SUS5\_memory \label{SUS5}\\
             6     & Frequency of thinking that user is in VE.                             & SUS6\_frequency present \label{SUS6}\\ \hline       
        \end{tabular}
        \caption{Overview of SUS questions examined in the study~\cite{slater_virtual_2000,usoh_using_2000}. The corresponding tag of each question will be used in the paper as a reference for that question.}
        \label{tab:questionnaire_sus}
        \Description{SUS questionnaire items. The columns of the table from left to right are item number, summary and a tag to reference it in later interpretations.}
    \end{table}
    
    \begin{table}[h]
        \centering
        \begin{tabular}{| l p{8.5cm} l |}\hline
            \multicolumn{3}{|l|}{\textbf{Presence Questionnaire by Witmer and Singer (PQ, 1998)~\cite{witmer1998measuring}}}\\
            Question & Description                                                          & Tag \\ \hline
            1     & Rate the ability to control events. (INV)                                   & \hypertarget{PQ1}{PQ1\_control events}\\      
            2     & Rate responsiveness to users' actions. (INV)                               & \hypertarget{PQ2}{PQ2\_responsiveness}\\               
            3     & How natural was the interaction? (NAT)                                     & \hypertarget{PQ3}{PQ3\_natural interaction}\\    
            4     & Level involvement through visual aspects. (INV)                            & \hypertarget{PQ4}{PQ4\_visual involvement}\\              
            5     & How natural was the control mechanism? (NAT)                               & \hypertarget{PQ5}{PQ5\_natural control}\\   
            6     & How compelling was sense of moving objects? (INV)                          & \hypertarget{PQ6}{PQ6\_compelling objects}\\     
            7     & Consistency with real world experiences? (NAT)                             & \hypertarget{PQ7}{PQ7\_consistency}\\          
            8     & Ability to anticipate what will happen based on users' actions. (INV)      & \hypertarget{PQ8}{PQ8\_anticipation}\\               
            9     & Ability to visually inspect the environment. (INV)                         & \hypertarget{PQ9}{PQ9\_visual exploration}\\      
            10    & How compelling was sense of moving around in the VE? (INV)                 & \hypertarget{PQ10}{PQ10\_compelling navigation}\\        
            11    & Ability to closely examine objects. (RES)                                  & \hypertarget{PQ11}{PQ11\_examination detail}\\      
            12    & Ability to investigate objects from multiple viewpoints. (RES)             & \hypertarget{PQ12}{PQ12\_examination viewpoints}\\ 
            13    & Involvement by VE experience. (INV)                                        & \hypertarget{PQ13}{PQ13\_involvement}\\
            14    & How much delay was between action and outcome? (INV)                       & \hypertarget{PQ14}{PQ14\_delay}\\
            15    & Speed of adjusting to the VE. (INV)                                        & \hypertarget{PQ15}{PQ15\_adjustment}\\          
            16    & Proficiency of movement and interaction in the VE. (INV)                   & \hypertarget{PQ16}{PQ16\_proficiency}\\           
            17    & Extent of interference by display quality. (IQ)                            & \hypertarget{PQ17}{PQ17\_display quality}\\         
            18    & How much did the control devices interfere the experience? (IQ)            & \hypertarget{PQ18}{PQ18\_interference control}\\             
            19    & Ability to concentrate on task rather on control mechanism. (IQ)           & \hypertarget{PQ19}{PQ19\_task focus}\\ \hline
        \end{tabular}
        \caption{Overview of the PQ questions examined in the study~\cite{witmer1998measuring}. The presence dimension of each item is given in parentheses after each question. The dimensions included are Involvement (INV, 11 items), Naturalness (NAT, 3 items), Resolution (RES, 2 items), and Interface Quality (IQ, 3 items). The corresponding tag of each question will be used in the paper as a reference for the respective question.}
        \Description{The PQ items. The columns of the table from left to right are item number, summary and a tag to reference it in later interpretations.}
        \label{tab:questionnaire_pq}
    \end{table}

    \clearpage

    \begin{table}[h]
        \centering
        \begin{tabular}{| l p{8cm} l |}\hline
            \multicolumn{3}{|l|}{\textbf{The IGroup Presence Questionnaire by Schubert (IPQ, 2001)~\cite{schubert_experience_2001,schubert_sense_2003}}}\\
            Question & Description                                                          & Tag \\ \hline
             1     & Rate the sense of being there. (PRES)                                       & \hypertarget{IPQ1}{IPQ1\_presence overall}\\
             2     & Feeling of being surrounded by the VE. (SP)                                 & \hypertarget{IPQ2}{IPQ2\_surrounding}\\
             3     & Extent of not feeling present in the VE. (SP)                               & \hypertarget{IPQ3}{IPQ3\_images}\\
             4     & Sense of acting in the VE instead from outside. (SP)                        & \hypertarget{IPQ4}{IPQ4\_no presence}\\
             5     & Feeling of being present in the VE. (SP)                                    & \hypertarget{IPQ5}{IPQ5\_acting inside}\\
             6     & Awareness of real world when navigating through VE. (SP)                    & \hypertarget{IPQ6}{IPQ6\_presence}\\
             7     & Not being aware of the real environment. (INV)                              & \hypertarget{IPQ7}{IPQ7\_awareness navigation}\\
             8     & Extent of not being aware of the real environment. (INV)                    & \hypertarget{IPQ8}{IPQ8\_awareness realworld}\\
             9     & How much did the user still pay attention to the real world? (INV)          & \hypertarget{IPQ9}{IPQ9\_attention realworld}\\
             10    & Extent of being captivated by the VE. (INV)                                 & \hypertarget{IPQ10}{IPQ10\_captivated}\\
             11    & Extent to which VE seemed to be real. (REAL)                                & \hypertarget{IPQ11}{IPQ11\_realism1}\\
             12    & Consistency of experience with real-world experience. (REAL)                & \hypertarget{IPQ12}{IPQ12\_consistency}\\
             13    & How real seemed VE? (REAL)                                                  & \hypertarget{IPQ13}{IPQ13\_realism2}\\
             14    & Was VE more realistic than real world? (REAL)                               & \hypertarget{IPQ14}{IPQ14\_realism dominance}\\ \hline
        \end{tabular}
        \caption{Overview of IPQ questions investigated in the study~\cite{schubert_experience_2001,schubert_sense_2003}. The presence dimension of each item is given in parentheses after each question. In addition to presence in general (PRES), the dimensions included are spatial presence (SP, 5 items), involvement (INV, 4 items), and realness (REAL, 4 items). The corresponding tag of each question will be used in the paper as a reference for the respective question.}
        \Description{The IPQ items. The columns of the table from left to right are item number, summary and a tag to reference it in later interpretations.}
        \label{tab:questionnaire_ipq}
    \end{table}

    \begin{table}[h]
        \centering
        \begin{tabular}{| l p{8cm} l |}\hline
            \multicolumn{3}{|l|}{\textbf{The single item presence scale by Bouchard (SIP, 2004)~\cite{bouchard_reliability_2004}}}\\
            Question & Description                                                          & Tag \\ \hline
             1     & To what extent do you feel present in the virtual environment, as if you were really there?                                        & SIP\_presence overall\label{SIP}\\ \hline    
        \end{tabular}
        \caption{Overview of the single-item presence scale by Bouchard (2004) studied in~\cite{bouchard_reliability_2004}. The corresponding tag will be used in the paper as a reference for the respective question.}
        \Description{Overview of the SIP questionnaire. The columns from left to right are item number, summary and a tag to reference it in later interpretations.}
        \label{tab:questionnaire_bouchard}
    \end{table}  

\clearpage

\section{Descriptive Analysis}

\subsection{Score differences}

\begin{table}[h]
    \centering
    \begin{tabular}{|r|c|c|c|c|c|c|c|c|c|c|}\hline
         Questionnaire & \multicolumn{10}{c|}{SUS} \\
         Presence Dimension     & \multicolumn{2}{c}{PI}&  \multicolumn{2}{c}{INV} &  \multicolumn{2}{c}{PSI} &  \multicolumn{2}{c}{SOC} &  \multicolumn{2}{c|}{EMB}\\
         Dimension Intensity    &  POS      &   NEG     &   POS     &   NEG     &   POS     &   NEG     &   POS     &   NEG     &   POS     &   NEG \\\hline
         Average Score ($M$)    &  5.27     &   4.48    &   4.13    &   3.82    &   4.68    &   4.35    &   4.52    &   4.29    &   4.46    &   4.32\\
         Median Score ($Md$)      &  5.33     &   4.50    &   4.33    &   3.83    &   4.58    &   4.25    &   4.50    &   4.33    &   4.42    &   4.33\\
         Standard Deviation ($\sigma$) & 0.88     &   1.30    &   1.26    &   1.21    &   1.10    &   1.17    &   1.15    &   1.24    &   1.16    &   1.15\\\hline
    \end{tabular}
    
    \begin{tabular}{|r|c|c|c|c|c|c|c|c|c|c|}\hline
         Questionnaire & \multicolumn{10}{c|}{PQ} \\
         Presence Dimension     & \multicolumn{2}{c}{PI}&  \multicolumn{2}{c}{INV} &  \multicolumn{2}{c}{PSI} &  \multicolumn{2}{c}{SOC} &  \multicolumn{2}{c|}{EMB}\\
         Dimension Intensity    &  POS      &   NEG     &   POS     &   NEG     &   POS     &   NEG     &   POS     &   NEG     &   POS     &   NEG \\\hline
         Average Score ($M$)    &  5.69     &   4.48    &   4.66    &   4.44    &   4.39    &   4.16    &   4.55    &   4.96    &   5.48    &   5.45\\
         Median Score ($Md$)      &  5.68     &   5.55    &   4.42    &   4.47    &   5.00    &   4.21    &   5.00    &   5.05    &   5.53    &   5.45\\
         Standard Deviation ($\sigma$) & 0.83     &   0.95    &   1.03    &   1.10    &   2.02    &   1.11    &   1.98    &   0.96    &   0.68    &   0.70\\\hline
    \end{tabular}

    \begin{tabular}{|r|c|c|c|c|c|c|c|c|c|c|}\hline
         Questionnaire & \multicolumn{10}{c|}{IPQ} \\
         Presence Dimension     & \multicolumn{2}{c}{PI}&  \multicolumn{2}{c}{INV} &  \multicolumn{2}{c}{PSI} &  \multicolumn{2}{c}{SOC} &  \multicolumn{2}{c|}{EMB}\\
         Dimension Intensity    &  POS      &   NEG     &   POS     &   NEG     &   POS     &   NEG     &   POS     &   NEG     &   POS     &   NEG \\\hline
         Average Score ($M$)    &  5.21     &   4.27    &   4.12    &   3.80    &   4.81    &   4.41    &   4.54    &   4.34    &   4.59    &   4.56\\
         Median Score ($Md$)      &  5.21     &   4.32    &   4.18    &   3.86    &   4.71    &   4.39    &   4.64    &   4.36    &   4.50    &   4.64\\
         Standard Deviation ($\sigma$) & 0.89     &   1.12    &   1.15    &   1.23    &   1.06    &   1.13    &   1.11    &   1.22    &   1.02    &   1.07\\\hline
    \end{tabular}
        
    \begin{tabular}{|r|c|c|c|c|c|c|c|c|c|c|}\hline
         Questionnaire & \multicolumn{10}{c|}{SIP} \\
         Presence Dimension     & \multicolumn{2}{c}{PI}&  \multicolumn{2}{c}{INV} &  \multicolumn{2}{c}{PSI} &  \multicolumn{2}{c}{SOC} &  \multicolumn{2}{c|}{EMB}\\
         Dimension Intensity    &  POS      &   NEG     &   POS     &   NEG     &   POS     &   NEG     &   POS     &   NEG     &   POS     &   NEG \\\hline
         Average Score ($M$)    &  5.68     &   4.37    &   4.21    &   3.72    &   4.90    &   4.45    &   4.31    &   4.79    &   4.45    &   4.46\\
         Median Score ($Md$)      &  5.60     &   4.20    &   3.85    &   3.85    &   4.90    &   4.90    &   4.90    &   4.90    &   4.90    &   4.90\\
         Standard Deviation ($\sigma$) & 0.93     &   1.63    &   1.80    &   1.73    &   1.63    &   1.79    &   1.52    &   1.62    &   1.56    &   1.47\\\hline
    \end{tabular}
    \caption{Mean, median, and standard deviation of all questionnaire scores split by scenario and condition. Visualization in \autoref{fig:scores_perScenario}.}
    \Description{Results of the descriptive analysis for research question 1. It shows the average scores, the median of the scores and standard deviation per questionnaire, scenario and embodiment.}
    \label{tab:scores_overview}
\end{table}

\begin{table}[h]
    \centering
    \begin{tabular}{|r|c|c|c|}\hline
         \textbf{Scenario} & \textbf{Questionnaire} & \textbf{Item} & \textbf{Significance}\\\hline
         PI & SUS     &   1-3,5,6    &   $p<.004$**\\
         PI & PQ  & 3,4,6,7,10,11,13,14,17 &  $p<0.017$*\\
         PI & IPQ & 1-14 & $p<0.02$*\\
         PI & SIP & 1 & $p<.001$***\\\hline
         INV &SUS&   2,4    &   $p<.042$*\\
         INV &PQ& 3,6,8,13 &  $p<0.038$*\\
         INV &IPQ& 2,4,5,8-10 & $p<0.046$*\\
         INV &SIP& 1 & $p<.007$**\\\hline
         PSI &SUS&1,2,5&$p<.005$**\\
         PSI &PQ & 1-11,13-16,18,19 &  $p<0.03$*\\
         PSI &IPQ& 1,2,5,6,10-13 & $p<0.05$*\\
         PSI &SIP& 1 & $p<.011$*\\\hline
         SOC &SUS& 2,5&$p<.037$*\\
         SOC &PQ& 3,5-9,14-16&  $p<0.05$*\\
         SOC &IPQ& 5,12 & $p<0.035$*\\
         SOC &SIP& 1 & $p<.019$*\\\hline
         EMB &SUS&4,6& $p<.0036$**\\\hline
    \end{tabular}
    \caption{The table shows all items where the \emph{paired-samples Wilcoxon signed rank test} for non-parametric data reported significant differences between the scores of positive and negative condition ($p<0.05$, $N=50$).}
    \Description{Summary of questionnaire item-dimension combinations that showed significant differences between negative and positive condition. From left to right, dimension, questionnaire, items and significance level are listed.}
    \label{tab:item_analysis_per_scenario}
\end{table}

\begin{table}[h]
    \centering
    \begin{tabular}{|r|c|c|c|c|c|c|c|}\hline
            & \multicolumn{5}{c|}{\textbf{Presence dimension}} \\
            & \textbf{PI} & \textbf{INV} & \textbf{PSI} & \textbf{SOC} & \textbf{EMB}\\\hline 
         \multirow{2}{*}{\textbf{SUS}} &  $M=0.77, Md=0.67$    &   $M=0.32, Md=0.33$   &   $M=0.33, Md=0.17$    &   $M=0.23, Md=0.17$ & $M=0.14, Md=0.17$ \\
         & $\sigma = 1.43$ & $\sigma = 1.52$ & $\sigma = 1.22$ & $\sigma = 1.49$ & $\sigma = 1.49$ \\ \hline
         \multirow{2}{*}{\textbf{PQ}} &  $M=0.31, Md=0.34$     &   $M=0.22, Md=0.16$   &   $M=1.17, Md=1.00$    &   $M=0.38, Md=0.3$ & $M=-0.01, Md=-.11$ \\  
         & $\sigma = 1.14$ & $\sigma = 1.30$ & $\sigma = 1.48$ & $\sigma = 1.36$ & $\sigma = 1.27$ \\ \hline    
         \multirow{2}{*}{\textbf{IPQ}} &  $M=0.88, Md=0.89$     &   $M=0.29, Md=0.14$   &   $M=0.4, Md=0.32$    &   $M=0.22, Md=0.18$ & $M=0.05, Md=0.07$ \\ 
         & $\sigma = 1.09$ & $\sigma = 1.09$ & $\sigma = 1.05$ & $\sigma = 1.34$ & $\sigma = 1.28$ \\ \hline
         \multirow{2}{*}{\textbf{SIP}} &  $M=1.15, Md=1.40$     &   $M=0.55, Md=0.70$   &   $M=0.47, Md=0.00$    &   $M=0.45, Md=0.00$ & $M=0.09, Md=0.0$ \\
         & $\sigma = 1.6$ & $\sigma = 1.87 $ & $\sigma = 1.83 $ & $\sigma = 1.48$ & $\sigma = 1.67$ \\\hline
    \end{tabular}
    \caption{Mean and median of the differences of presence scores (score of positive - negative condition) from the questionnaires split by conditions. Visualization in figure \ref{fig:score differences per questionnaire and scenario}.}
    \Description{Results of the descriptive analysis for research question 1. It includes the average score differences, the median of the score differences and standard deviation per questionnaire and scenario.}
    \label{tab:approval_overview}
\end{table}

\clearpage

\subsection{Sensitivity of Questionnaire Items}

    \begin{figure}[h]
        \centering
            \includegraphics[width=\textwidth]{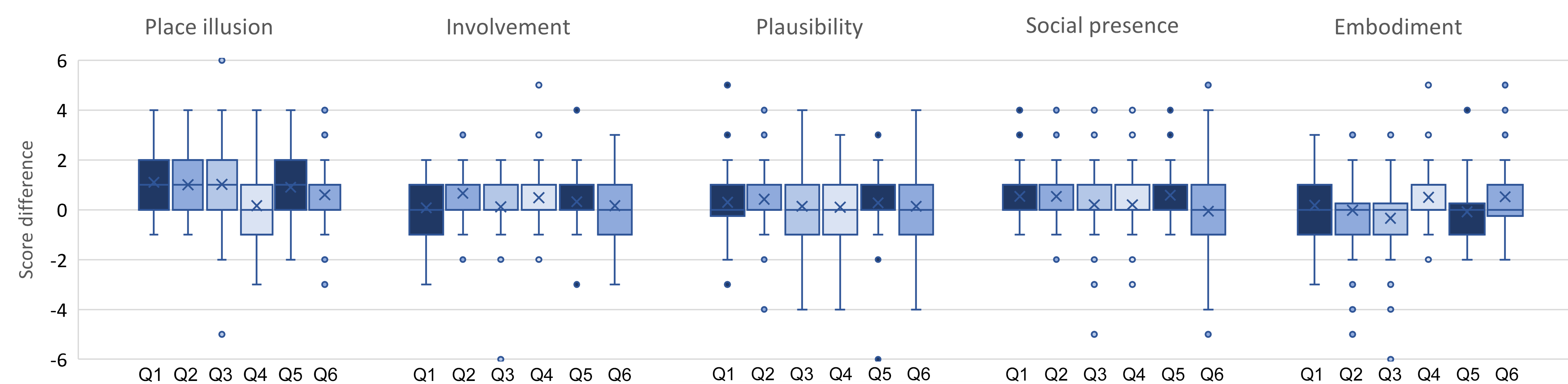}
            \caption{SUS score differences between positive and negative condition grouped by scenario.}
            \Description{Boxplot chart showing score differences of the SUS for the five dimensions: Place Illusion, Involvement, Plausibility, Social Presence, and Embodiment. The y-axis indicates score variations ranging from -6 to 6, with each questionnaire item represented by a distinct boxplot grouped by dimension. Overall, the score variations were predominantly positive, with means ranging from -0.34 (question 3 for changes in the embodiment dimension) to 1.02 (question 3 for changes in the place illusion dimension). The interquartile ranges exhibited variability across factors, with the Place Illusion and Plausibility factors demonstrating the widest spread (IQR = 2 for three or more items). With regard to the score differences associated with the plausibility scenario, opinions concerning the differences appear to be divided, particularly for questions 3, 4, and 6, as score differences were prevalent with means approximating 0.1. In contrast, involvement, social presence, and embodiment exhibited a more concentrated distribution, with IQRs ranging around 1. The sixth question exhibited a more dispersed distribution for the changes in involvement, plausibility, and social presence dimensions, while score differences varied less for the other dimensions. Outliers were identified for all dimensions, with the fewest occurring for the place illusion dimension. }
            \label{fig:SUS_sensitivity_items}
        \end{figure} 

    \begin{figure}[h]
        \centering
            \includegraphics[width=\textwidth]{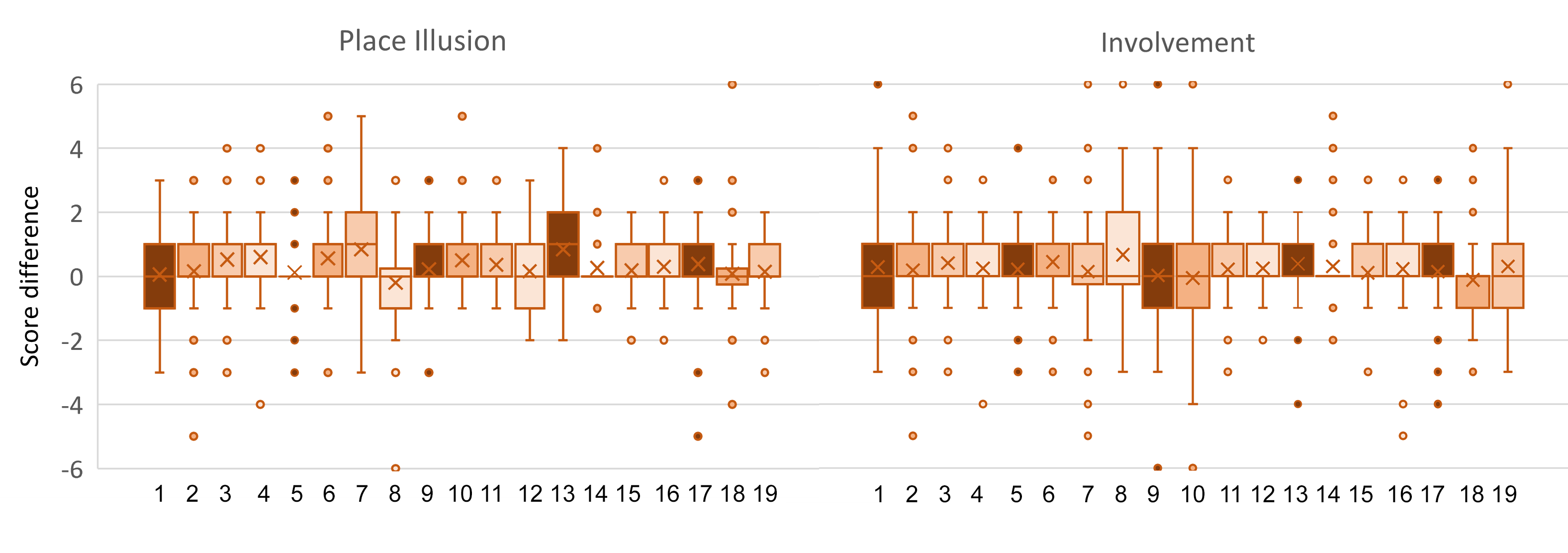}
            \caption{PQ score differences between positive and negative condition of the place illusion and involvement scenario.}
            \Description{Boxplot chart of two dimensions showing score differences of the PQ for the place illusion  scenario on the left and the involvement scenario on the right. The y-axis denotes score variations ranging from -6 to 6, with each question on the x-axis (Q1-Q19) represented by a distinct boxplot. The mean is depicted as a cross, while the median is represented by a horizontal line. In the place illusion section, score variations predominantly exhibit slight positivity, with means ranging from -0.2 to 0.84. The highest score disparities were observed in questions 7 and 13 (both M = 0.84). Items 1, 7, 12, and 13 exhibited a wider interquartile range (IQR = 2). Questions 5 and 14 exhibited a concentrated distribution with few outliers ( IQR = 0), suggesting consistency across participants. The involvement section displays analogous distributions, with score differences falling within an IQR of 1 for the majority of questions, exhibiting slightly positive means ranging from -0.12 to 0.66. However, certain questions exhibit a more extensive spread, with an IQR of approximately 2, including questions 1, 8-10, and 19.In summary, both factors demonstrate variations in score differences, with the presence of multiple outliers across items.}
            \label{fig:PQ_sensitivity_items1}
        \end{figure} 

    \begin{figure}[h]
        \centering
            \includegraphics[width=\textwidth]{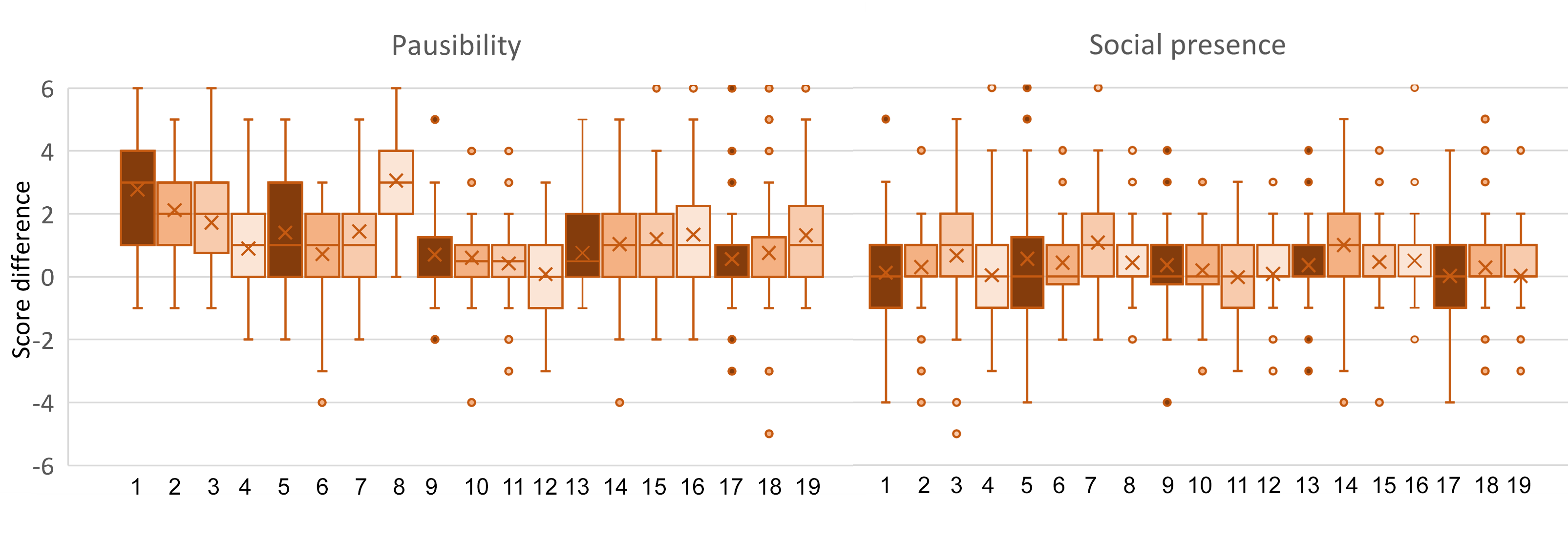}
            \caption{PQ score differences between positive and negative condition of the plausibility and social presence scenario.}
            \Description{Boxplot chart of two dimensions showing score differences of the PQ for the Plausibility  scenario on the left and the social presence scenario on the right. The y-axis denotes score variations ranging from -6 to 6, with each question on the x-axis (Q1-Q19) represented by a distinct boxplot. The mean is depicted as a cross, while the median is represented by a horizontal line.In the Plausibility section, score differences predominantly exhibit a positive trend, with means ranging from 0.08 to 3.06. The most significant score differences are observed in questions 1 (M = 2.78) and 8 (M = 3.06). Several items demonstrate a wider interquartile range, such as questions 1 and 5, with an IQR of 3, suggesting mixed opinions on the differences between positive and negative conditions. Conversely, questions 9–11 and 17–18 exhibited a more concentrated distribution (IQR between 1 and 1.25), suggesting consistency across participants. The social presence section demonstrates smaller score variations, with means ranging from -0.02 to 1. While the distribution of score differences is largely comparable for the majority of the questions, with an IQR of approximately 1, questions 1, 3-5, 7, 11, 14, and 17 exhibit a more extensive distribution with an IQR of 2. Overall, both factors demonstrate variations in score differences, with the presence of multiple outliers across items.}
            \label{fig:PQ_sensitivity_items2}
        \end{figure} 

    \begin{figure}[h]
        \centering
            \includegraphics[width=0.5\textwidth]{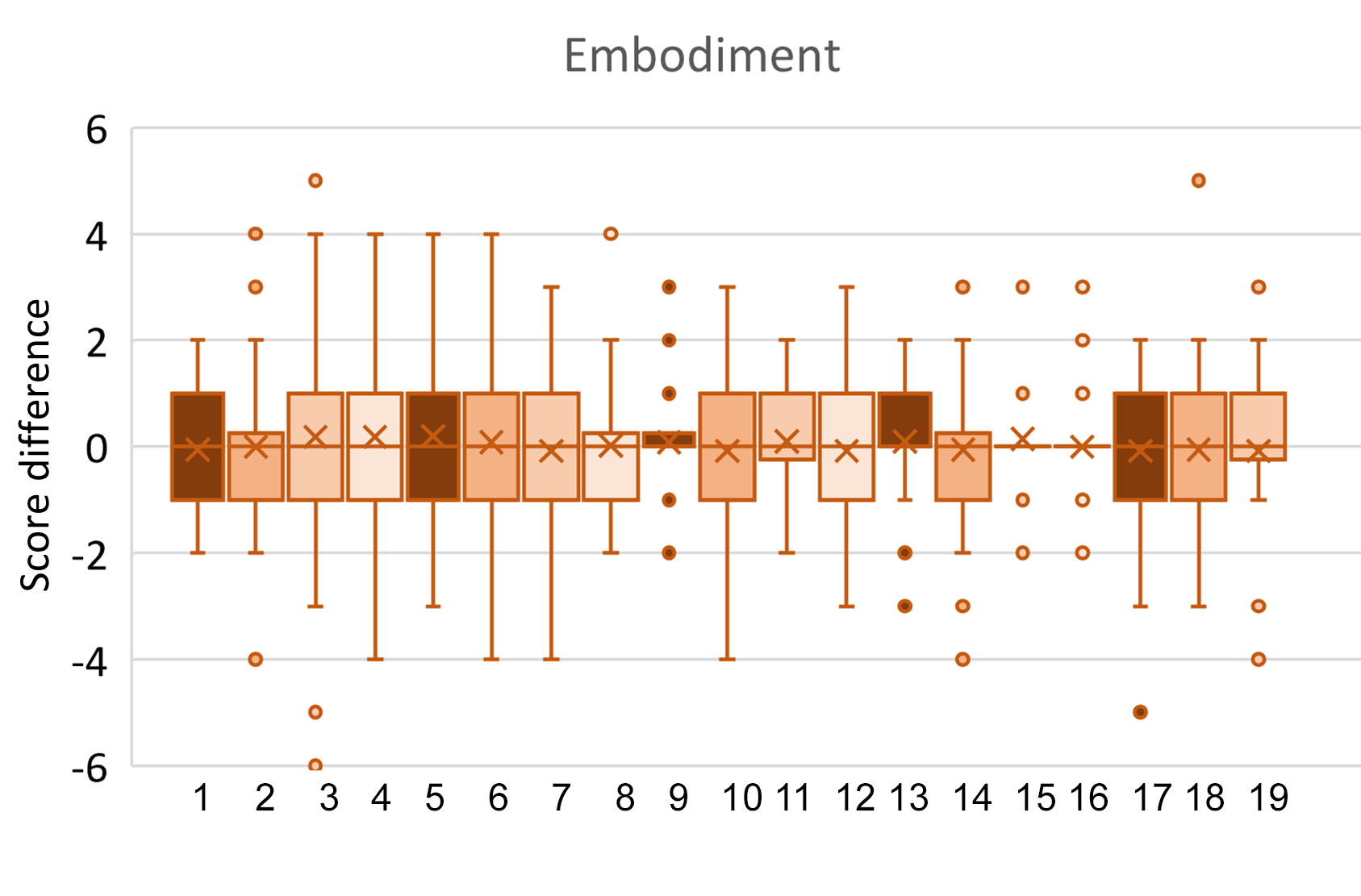}
            \caption{PQ score differences between positive and negative condition of the embodiment scenario.}
                \Description{Boxplot chart showing score differences of the PQ for the changes in the embodiment dimension. The y-axis represents score differences ranging from minus 6 to 6, while each question on the x-axis (Q1-Q19) is represented by a separate boxplot. The score differences are mostly around 0 with means between minus 0,08 and 0,2 and IQRs between 0 and 2. The least uncertainty is observable for question 9 (M= 0,008 and IQR= 0,25), for question 15 (M= 0,14 and IQR= 0) and for question 16 (M= 0 and IQR= 0). The highest uncertainty is observable for questions 3-7 ranging from minus 4 to 4 with IQRs of 2. }
            \label{fig:PQ_sensitivity_items3}
        \end{figure}     

    \begin{figure}[h]
        \centering
            \includegraphics[width=\textwidth]{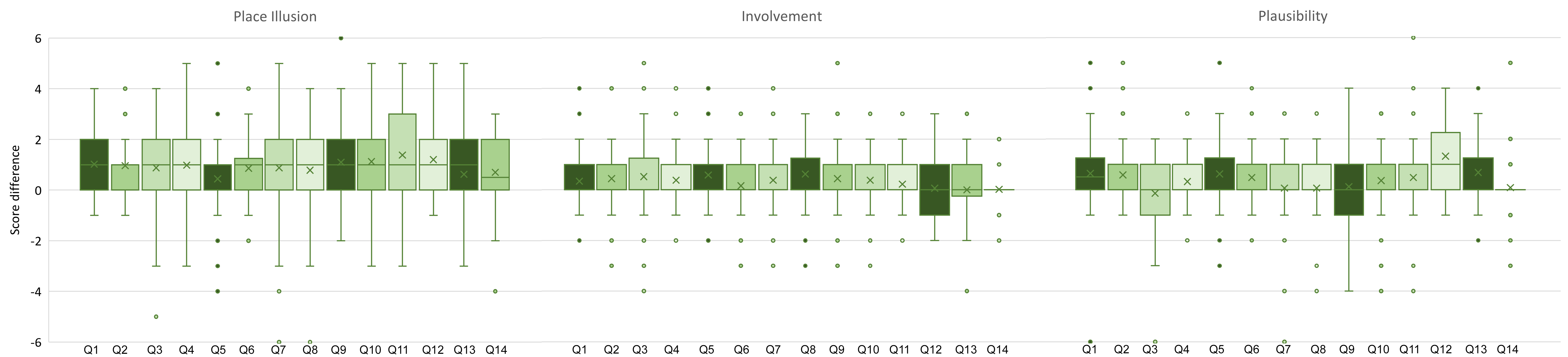}
            \caption{IPQ score differences between positive and negative condition of the place illusion, involvement, and plausibility scenario.}
            \Description{Three Boxplot charts showing IPQ score differences between positive and negative conditions for place illusion on the left, involvement in the middle and plausibility on the right. The x-axis of the figure represents the individual questionnaire items (Q1–Q14) of the IPQ, while the y-axis represents score differences ranging from -6 to 6. Each item is represented by a separate boxplot, with interquartile ranges, means, and outliers displayed. The variability is observed to be the highest for the score differences in the place illusion scenario, ranging from -3 to 5 and IQRs between 1 and 3. The lowest observed level of uncertainty was identified in question 5, with a mean of 0.44 and an IQR of 1. In contrast, question 11 exhibited the greatest uncertainty, with a mean of 1.38 and an IQR of 3. The score differences for the involvement and plausibility dimensions are predominantly within the same range, with an IQR of approximately 1 and means ranging from 0 to 0.62 for the changes in the involvement dimension and from -0.14 to 1.32 for the changes in the plausibility dimension. The most prevalent distribution concerning the score differences of the involvement dimension is evident in question 12, with a mean of 0.06 and an IQR of 2. With regard to the plausibility score differences, the greatest uncertainty was observed in question 3 (M= -0.14 and IQR= 2), question 9 (M= 0.12 and IQR = 2), and question 12 (M = 1.32 and IQR = 2.25). However, the latter two dimensions exhibited a greater number of outliers compared to the place illusion scenario.}
            \label{fig:IPQ_sensitivity_items1}
        \end{figure} 

    \begin{figure}[h]
        \centering
            \includegraphics[width=0.7\textwidth]{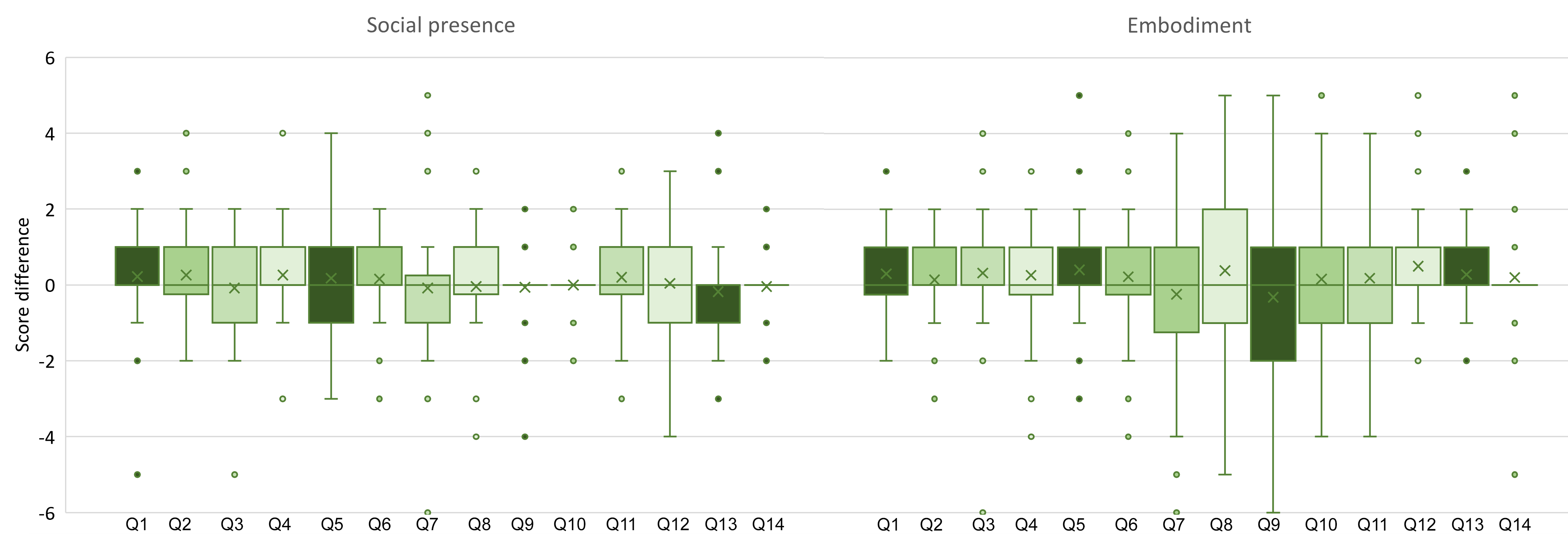}
            \caption{IPQ score differences between positive and negative condition of the social presence and embodiment scenario.}
            \Description{Two Boxplot charts showing IPQ score differences of positive and negative conditions for social presence on the left and embodiment dimension on the right. The x-axis represents individual questionnaire items (Q1–Q14) of the IPQ, while the y-axis represents score differences ranging from minus 6 to 6. Each item has a separate boxplot, with interquartile ranges, means, and outliers displayed. Variability differs across items. The IQRs for the social presence dimension are smaller ranging from 0-2 while the IQRs of the embodiment dimension are between 0 and 3, with some showing wider distributions and more outliers.}
            \label{fig:IPQ_sensitivity_items2}
        \end{figure} 

    \begin{figure}[h]
        \centering
            \includegraphics[width=0.4\textwidth]{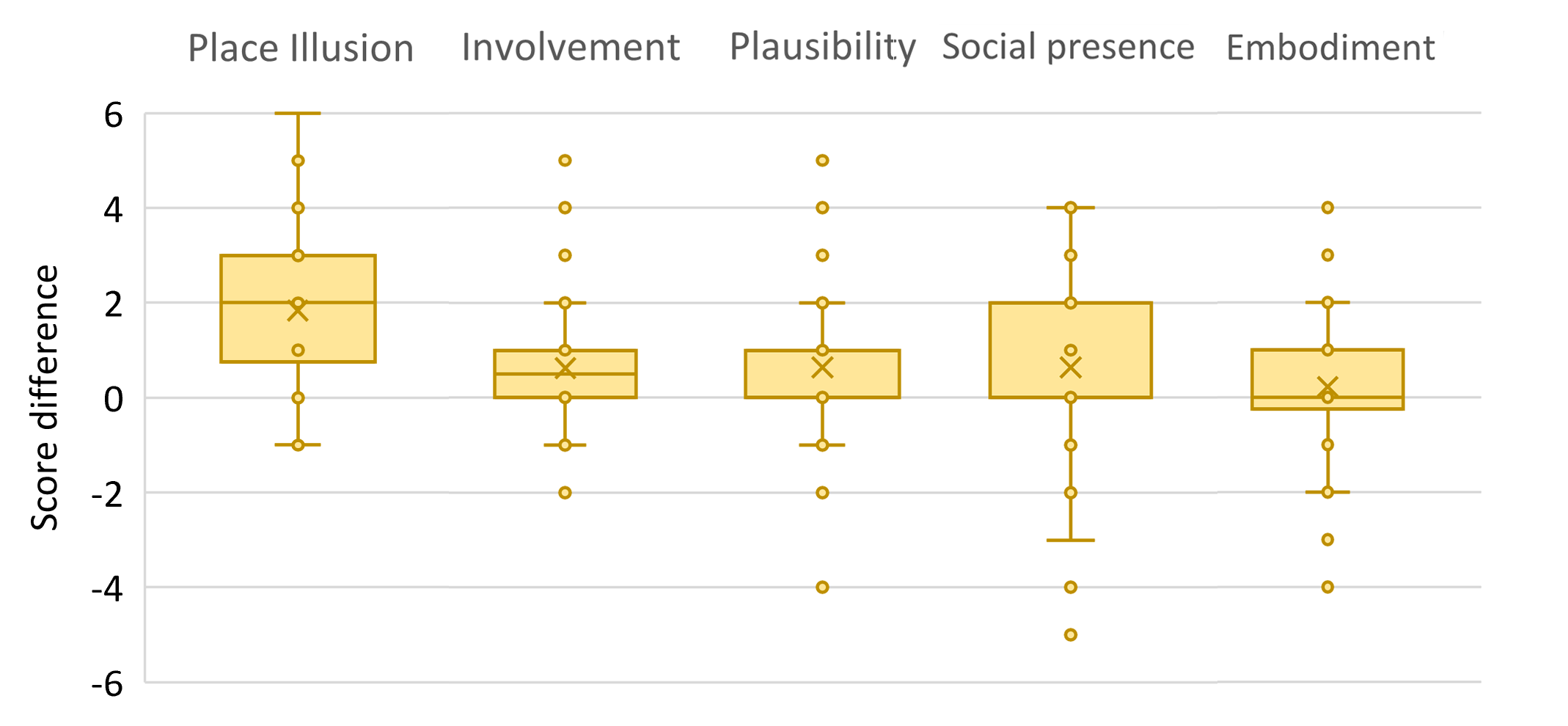}
            \caption{SIP score differences between positive and negative condition grouped by scenario.}
            \Description{Boxplot chart showing score differences of the single item questionnaire by Bouchard across five factors: Place Illusion, Involvement, Plausibility, Social Presence, and Embodiment. The y-axis of the chart represents score differences ranging from negative 6 to 6, with each dimension represented by a separate boxplot. The overall score differences were predominantly positive, with means ranging from 0.22 (embodiment) to 1.96 (place illusion). The interquartile range (IQR) varied across factors, with Place Illusion and social presence showing the widest spread ( IQR = 2) and involvement, plausibility, and embodiment around 1. The mean score difference ranged from 0.2 for changes of the embodiment dimension to 2 for the place illusion scenario, and outliers were present for all dimensions.}
            \label{fig:SIP_sensitivity_items}
        \end{figure} 

\clearpage

\subsection{Approval to Score Differences}

\begin{table}[h]
    \centering
    \begin{tabular}{|r|c|c|c|c|}\hline
         \textbf{Questionnaire} & \textbf{SUS} & \textbf{PQ} & \textbf{IPQ} & \textbf{SIP}\\\hline
         \multirow{2}{*}{\textbf{Score Difference}} & $M = 0.36, Md = 0.33$  &  $M = 0.43, Md = 0.37$  &  $M = 0.38, Md = 0.27$ &  $M = 0.57, Md = 0.7$\\
         & $\sigma = 0.88$ & $\sigma = 0.73$ & $\sigma = 0.77$ & $\sigma = 1.25$\\ \hline            
         \multirow{2}{*}{\textbf{Approval}} &  $M = -0.50, Md = -1.00$  &  $M = -0.40, Md = -1.00$ &  $M = -0.49, Md = 0.00$ & $M = -0.26, Md = 0.00$ \\
         & $\sigma = 1.48$ & $\sigma = 1.39$ & $\sigma = 1.21$ & $\sigma = 1.73$\\ \hline    
    \end{tabular}
    \caption{The table shows the average, median, and standard deviation of the score difference between the two conditions, as well as the mean, median, and standard deviation of participants' agreement with this difference and the columns representing the four questionnaires. For agreement with the difference, 0 is to be interpreted as the target value, as it means that the participants agree with the score difference. Positive values mean that the score is higher than anticipated, negative values that the score is lower than expected.}
    \Description{Results of the descriptive analysis for research question 2 showing the average score differences, the median of the score differences and standard deviation in comparison to the corresponding approval per questionnaire and scenario.}
    \label{tab:scores_approval}
\end{table}

\begin{table}[h]
    \centering
    \begin{tabular}{|r|c|c|c|c|c|c|c|}\hline
            & \multicolumn{5}{c|}{Presence dimension} \\
            & \textbf{PI} & \textbf{INV} & \textbf{PSI} & \textbf{SOC} & \textbf{EMB}\\\hline 
         \multirow{3}{*}{\textbf{SUS}} &  $M=-0.52$    &   $M=-0.06$   &   $M=-1.04$    &   $M=-0.78$ & $M=-0.10$ \\
          &  $Md=-1.00$    &   $Md=0.00$   &  $Md=-1.00$    &   $Md=-1.00$ & $Md=0.00$ \\
         & $\sigma = 1.43$ & $\sigma = 1.52$ & $\sigma = 1.22$ & $\sigma = 1.49$ & $\sigma = 1.49$ \\ \hline
         \multirow{3}{*}{\textbf{PQ}} &  $M=-1.16$     &   $M=-0.04$   &   $M=0.14$    &   $M=-0.52$ & $M=-0.49$ \\
          &  $Md=-1.00$     &   $Md=0.00$   &   $Md=0.00$    &   $Md=-0.50$ & $Md=-1.00$ \\  
         & $\sigma = 1.14$ & $\sigma = 1.30$ & $\sigma = 1.48$ & $\sigma = 1.36$ & $\sigma = 1.27$ \\ \hline    
         \multirow{3}{*}{\textbf{IPQ}} &  $M=-0.26$     &   $M=-0.18$   &   $M=-0.98$    &   $M=-0.46$ & $M=-0.42$ \\ 
         & $Md=0.00$     &   $Md=0.00$   &   $Md=-1.00$    &   $Md=-1.00$ & $Md=0.00$ \\ 
         & $\sigma = 1.09$ & $\sigma = 1.09$ & $\sigma = 1.05$ & $\sigma = 1.34$ & $\sigma = 1.28$ \\ \hline
         \multirow{3}{*}{\textbf{SIP}} &  $M=-0.20$     &   $M=0.26$   &   $M=-0.76$    &   $M=-0.20$ & $M=-0.38$ \\
         &  $Md=0.00$     &   $Md=0.00$   &   $Md=-1.00$    &   $Md=0.00$ & $Md=0.0$ \\
         & $\sigma = 1.6$ & $\sigma = 1.87 $ & $\sigma = 1.83 $ & $\sigma = 1.48$ & $\sigma = 1.67$ \\\hline
    \end{tabular}
    \caption{Mean, median, and standard deviation of the participants' approval to the differences of the presence scores (score of positive - negative condition) split by questionnaire and condition. Positive values mean the score difference is higher than anticipated, and negative values mean the score difference is lower than anticipated. Visualization in figure \ref{fig:presence_approval_scoredifferences}.}
    \Description{Results of the descriptive analysis for research question 2. It includes the average difference approval, the median of the difference approval, and its standard deviation.}
    \label{tab:approval_overview_conditions}
\end{table}

\clearpage

\section{Statistical Test Results}
\subsection{Score Differences per Questionnaire}
\begin{table}[ht]
        \centering
        \caption{Results of the related-samples Wilcoxon signed-rank test comparing the score of positive and negative condition.}
        \Description{Results of the Wilcoxon signed-rank test for the score differences. From left to right, dimension, questionnaire, Z-statistic, p-value, 95-percent confidence interval and effect size are listed.}
        \begin{tabular}{l l c c c c}
        \hline
        \textbf{Scenario} & \textbf{Questionnaire} & \textbf{Statistic (z)} & \textbf{p-value} & \textbf{CI} & \textbf{Effect Size (r)} \\ \hline
        Place Illusion (PI) & SUS & $5.095$ & $<0.001$ & $[0.5, 1.0]$ & $0.72$ \\ 
        & PQ  & $4.04$  & $<0.001$ & $[0.177, 0.435]$ & $0.57$ \\ 
        & IPQ & $5.797$ & $<0.001$ & $[0.785, 1.225]$ & $0.82$ \\ 
        & SIP & $5.33$  & $<0.001$ & $[1.050, 1.750]$ & $0.75$ \\ \hline
        Involvement (INV)   & SUS & $2.868$ & $0.004$  & $[0.083, 0.5]$ & $0.41$ \\ 
        & PQ  & $2.542$ & $0.011$  & $[0.056, 0.318]$ & $0.36$ \\ 
        & IPQ & $3.142$ & $0.002$  & $[0.151, 0.592]$ & $0.44$ \\ 
        & SIP & $2.687$ & $0.007$  & $[0.0, 0.7]$ & $0.38$ \\ \hline
        Plausibility (PSI)  & SUS & $2.714$ & $0.007$  & $[0.083, 0.5]$ & $0.38$ \\ 
        & PQ  & $5.652$ & $<0.001$ & $[0.903, 1.480]$ & $0.8$ \\ 
        & IPQ & $4.735$ & $0.001$  & $[0.385, 0.759]$ & $0.67$ \\ 
        & SIP & $2.554$ & $0.011$  & $[0.0, 0.7]$ & $0.36$ \\ \hline
        Social Presence (SOC) & SUS & $1.787$ & $0.074$  & $[0.0, 0.417]$ & $0.25$ \\ 
        & PQ  & $3.412$ & $<0.001$ & $[0.170, 0.556]$ & $0.48$ \\ 
        & IPQ & $1.829$ & $0.067$  & $[-0.024, 0.507]$ & $0.26$ \\ 
        & SIP & $2.342$ & $0.019$  & $[0.0, 0.7]$ & $0.33$ \\ \hline
        Embodiment (EMB)    & SUS & $-0.945$ & $0.345$  & $[-0.083, 0.333]$ & $0.13$ \\ 
        & PQ  & $1.849$  & $0.754$  & $[-0.01, 0.329]$ & $0.26$ \\ 
        & IPQ & $1.849$  & $0.065$  & $[-0.01, 0.329]$ & $0.26$ \\ 
        & SIP & $-0.945$ & $0.276$  & $[-0.083, 0.333]$ & $0.13$ \\ \hline
        \end{tabular}
            \label{tab:scoredifferencewilcoxon}
        \end{table}

\clearpage

\subsection{Score Differences per Item}

\begin{table}[h]
    \centering
    \begin{tabular}{|l|c|c|c|l|}\hline
         Question & Scenario & Sign. & Effect size & Tag\\\hline
        SUS 1 & PI & $<.001$ & 0.74 & \hyperlink{SUS1}{SUS1\_presence overall}\\
        SUS 1 & PSI & $.004$ & 0.40 &\\\hline
        SUS 2 & PI & $<.001$ & 0.65& \hyperlink{SUS2}{SUS2\_became reality}\\
        SUS 2 & INV & $<.001$ & 0.51 & \\
        SUS 2 & PSI & $.005$ & 0.39 & \\
        SUS 2 & SOC & $.003$ & 0.30 & \\\hline
        SUS 3 & PI & $<.001$ & 0.55 & \hyperlink{SUS3}{SUS3\_images}\\ \hline
        SUS 4 & INV & $.004$ & 0.29 & \hyperlink{SUS4}{SUS4\_dominant location}\\
        SUS 4 & EMB & $.019$ & 0.33 &\\ \hline
        SUS 5 & PI & $<.001$ & 0.52 &\hyperlink{SUS5}{SUS5\_memory}\\
        SUS 5 & PSI & $<.001$ & 0.51 &\\
        SUS 5 & SOC & $.037$ & 0.29 &\\\hline
        SUS 6 & PI & $.004$ & 0.41 & \hyperlink{SUS6}{SUS6\_frequency present}\\
        SUS 6 & EMB & $.036$ & 0.30 &\\\hline
        PQ 1 & PSI & $<.001$ & 0.81 & \hyperlink{PQ1}{PQ1\_control events}\\\hline
        PQ 2 & PSI & $<.001$ & 0.79 & \hyperlink{PQ2}{PQ2\_responsiveness}\\\hline
        PQ 3 & INV & $.024$ & 0.32 & \hyperlink{PQ3}{PQ3\_natural interaction}\\
        PQ 3 & PSI & $<.001$ & 0.74 &\\
        PQ 3 & SOC & $.001$ & 0.36 &\\\hline
        PQ 4 & PI & $.002$ & 0.45 & \hyperlink{PQ4}{PQ4\_visual involvement}\\
        PQ 4 & INV & $.072$ & 0.25 &\\
        PQ 4 & PSI & $<.001$ & 0.51 &\\\hline
        PQ 5 & PSI & $.002$ & 0.45 & \hyperlink{PQ5}{PQ5\_natural control}\\
        PQ 5 & SOC & $0.04$ & 0.29 &\\\hline
        PQ 6 & PI & $.008$ & 0.37 & \hyperlink{PQ6}{PQ6\_compelling objects}\\
        PQ 6 & INV & $.002$ & 0.32 &\\
        PQ 6 & PSI & $.004$ & 0.40 &\\
        PQ 6 & SOC & $<.05$ & 0.28 &\\\hline
        PQ 7 & PI & $<.001$ & 0.53 & \hyperlink{PQ7}{PQ7\_consistency}\\
        PQ 7 & PSI & $<.001$ & 0.70 &\\
        PQ 7 & SOC & $<.001$ & 0.59 &\\\hline
        PQ 8 & INV & $.016$ & 0.34 & \hyperlink{PQ8}{PQ8\_anticipation}\\
        PQ 8 & PSI & $<.001$ & 0.85 &\\
        PQ 8 & SOC & $.012$ & 0.35 &\\\hline
        PQ 9 & PSI & $<.001$ & 0.47 & \hyperlink{PQ9}{PQ9\_visual exploration}\\ \hline
        PQ 10 & PI & $.001$ & 0.45 & \hyperlink{PQ10}{PQ10\_compelling navigation}\\
        PQ 10 & PSI & $.002$ & 0.43 & \\ \hline
        PQ 11 & PI & $.005$ & 0.40 & \hyperlink{PQ11}{PQ11\_examination detail}\\
        PQ 11 & PSI & $.03$ & 0.31 & \\ \hline
        PQ 13 & PI & $<.001$ & 0.57 & \hyperlink{PQ13}{PQ13\_involvement}\\
        PQ 13 & INV & $.038$ & 0.29 & \\
        PQ 13 & PSI & $<.001$ & 0.52 & \\ \hline
        PQ 14 & PI & $.017$ & 0.34 & \hyperlink{PQ14}{PQ14\_delay}\\
        PQ 14 & PSI & $<.001$ & 0.50 & \\ 
        PQ 14 & SOC & $<.001$ & 0.52 & \\  \hline
        PQ 15 & PSI & $<.001$ & 0.59 & \hyperlink{PQ15}{PQ15\_adjustment}\\
        PQ 15 & SOC & $.007$ & 0.38 & \\  \hline
        PQ 16 & PSI & $<.001$ & 0.66 & \hyperlink{PQ16}{PQ16\_proficiency}\\
        PQ 16 & SOC & $.006$ & 0.38 & \\  \hline
        PQ 17 & PI & $.017$ & 0.34 & \hyperlink{PQ17}{PQ17\_display quality}\\
        PQ 17 & PSI & $.059$ & 0.27 & \\  \hline
        PQ 18 & PSI & $.010$ & 0.37 & \hyperlink{PQ18}{PQ18\_interface control}\\ \hline
        PQ 19 & PSI & $<.001$ & 0.64 & \hyperlink{PQ19}{PQ19\_task focus}\\ \hline
    \end{tabular}
    \caption{Results of the related-samples Wilcoxon signed-rank test for SUS and PQ showing a significant difference between the scores of the negative and positive condition per questionnaire item.}
    \Description{Significant results of the Wilcoxon signed-rank test for the score difference per questionnaire item and per dimension. The item, dimension, p-value, and effect size are listed from left to right.}
    \label{tab:item_analysis1}
\end{table}

\begin{table}[h]
    \centering
    \begin{tabular}{|l|c|c|c|l|}\hline
        IPQ 1 & PI & $<.001$ & 0.67 & \hyperlink{IPQ1}{IPQ1\_presence overall}\\
        IPQ 1 & PSI & $.001$ & 0.45 & \\ \hline
        IPQ 2 & PI & $<.001$ & 0.64 & \hyperlink{IPQ2}{IPQ2\_responsiveness}\\
        IPQ 2 & INV & $.021$ & 0.33 & \\ 
        IPQ 2 & PSI & $.002$ & 0.44 & \\ \hline
        IPQ 3 & PI & $<.001$ & 0.49 & \hyperlink{IPQ3}{IPQ3\_images}\\\hline
        IPQ 4 & PI & $<.001$ & 0.60 & \hyperlink{IPQ4}{IPQ4\_no presence}\\
        IPQ 4 & INV & $.031$ & 0.30 & \\\hline
        IPQ 5 & PI & $.020$ & 0.33 & \hyperlink{IPQ5}{IPQ5\_acting outside}\\
        IPQ 5 & INV & $.003$ & 0.43 & \\
        IPQ 5 & PSI & $.004$ & 0.41 & \\
        IPQ 5 & SOC & $.035$ & 0.30 & \\ \hline
        IPQ 6 & PI & $<.001$ & 0.58 & \hyperlink{IPQ6}{IPQ6\_presence}\\
        IPQ 6 & PSI & $.008$ & 0.38 & \\ \hline
        IPQ 7 & PSI & $<.001$ & 0.48 & \hyperlink{IPQ7}{IPQ7\_awareness navigation}\\ \hline
        IPQ 8 & PI & $<.001$ & 0.48 & \hyperlink{IPQ8}{IPQ8\_awareness realworld}\\
        IPQ 8 & INV & $.003$ & 0.30 & \\ \hline
        IPQ 9 & PI & $<.001$ & 0.55 & \hyperlink{IPQ9}{IPQ9\_attention realworld}\\
        IPQ 9 & INV & $.036$ & 0.30 & \\ \hline
        IPQ 10 & PI & $<.001$ & 0.63 & \hyperlink{IPQ10}{IPQ10\_captivated}\\
        IPQ 10 & INV & $.046$ & 0.28 & \\
        IPQ 10 & PSI & $.050$ & 0.28 & \\ \hline
        IPQ 11 & PI & $<.001$ & 0.66 & \hyperlink{IPQ11}{IPQ11\_realism1}\\
        IPQ 11 & PSI & $.044$ & 0.28 & \\ \hline
        IPQ 12 & PI & $<.001$ & 0.67 & \hyperlink{IPQ12}{IPQ12\_consistency}\\
        IPQ 12 & PSI & $<.001$ & 0.66 & \\ 
        IPQ 12 & SOC & $.026$ & 0.32 & \\\hline
        IPQ 13 & PI & $.014$ & 0.35 & \hyperlink{IPQ13}{IPQ13\_realism2}\\
        IPQ 13 & PSI & $.001$ & 0.45 &\\\hline
        IPQ 14 & PI & $<.001$ & 0.49 & \hyperlink{IPQ14}{IPQ14\_realism dominance} \\ \hline
        SIP & PI & $<.001$ & 0.75 & \hyperlink{SIP}{SIP\_presence overall}\\
        SIP & INV & $.007$ & 0.38 & \\
        SIP & PSI & $.011$ & 0.36 & \\
        SIP & SOC & $.019$ & 0.33 & \\ \hline
    \end{tabular}
    \caption{Results of the related-samples Wilcoxon signed-rank test for IPQ and SIP showing a significant difference between the scores of the negative and positive condition per questionnaire item.}
    \Description{Second part of Table 13 presenting an overview of the significant results of the Wilcoxon signed-rank test for score differences per questionnaire item and dimension. The item, dimension, p-value, and effect size are listed from left to right.}
    \label{tab:item_analysis2}
\end{table}

\clearpage

    \subsection{Analysis of Presence Dimensions Included in PQ and IPQ}
 
    \begin{table}[h]
        \centering
        \begin{tabular}{|l|c|c|}\hline
             \textbf{Scenario/ Dim}& \textbf{PQ: Involvement} & \textbf{PQ: Naturalness}\\
             Place illusion & (z=3.605, p<0.001, r=0.51, CI=[0.136,0.445]) & (z=3.484, p<0.001, r=0.49, CI=[0.333,0.833])\\ \hline
             Involvement & (z=3.527, p<0.001, r=0.50, CI=[0.136,0.409]) & (z=1.574, p=0.116, r=0.22, CI=[0.0,0.5])\\ \hline
             Plausibility & (z=6.107, p<0.001, r=0.86, CI=[1.091,1.636]) & (z=5.793, p<0.001, r=0.82, CI=[1.0,1.833])\\ \hline
             Social Presence & (z=3.530, p<0.001, r=0.50, CI=[0.136,0.5]) & (z=3.649, p<0.001, r=0.52, CI=[0.333,1.0])\\ \hline
             Embodiment & (z=0.19, p=0.849, r=0.03, CI=[-0.182,0.227])  & (z=0.234, p=0.815, r=0.03, CI=[-0.333,0.5])\\ \hline
             \textbf{Scenario/ Dim}& \textbf{PQ: Resolution} & \textbf{PQ: Interface quality}\\ 
             Place Illusion & (z=2.235, p=0.025, r=0.32, CI=[0.0,0.5]) & (z=1.591, p=0.112, r=0.23, CI=[0.0,0.333])\\ \hline
             Involvement & (z=1.990, p=0.047, r=0.28, CI=[0.0,0.5]) & (z=0.631, p=0.528, r=0.09, CI=[-0.167,0.333])\\ \hline
             Plausibility & (z=1.376, p=0.169, r=0.19, CI=[0.0,0.5]) & (z=3.967, p<0.001, r=0.56, CI=[0.333,1.167])\\ \hline
             Social Presence & (z=0.21, p=0.834, r=0.03, CI=[-0.25,0.25]) & (z=0.299, p=0.765, r=0.04, CI=[-0.167,0.333])\\ \hline
             Embodiment & (z=-2.192, p=0.028, r=0.31, CI=[-1.125,-0.125])  & (z=-1.182, p=0.237, r=0.17, CI=[-0.75,0.25])\\ \hline
             \textbf{Scenario/ Dim}& \textbf{IPQ: Spatial presence} & \textbf{IPQ: Involvement}\\
             Place Illusion & (z=4.718, p<0.001, r=0.67, CI=[0.5,1.1]) & (z=4.686, p<0.001, r=0.66, CI=[0.625,1.250])\\ \hline
             Involvement & (z=3.183, p=0.001, r=0.45, CI=[0.2,0.6]) & (z=3.123, p=0.002, r=0.44, CI=[0.125,0.75])\\ \hline
             Plausibility & (z=2.494, p=0.013, r=0.35, CI=[0.1,0.6]) & (z=1.647, p=0.1, r=0.23, CI=[0.0,0.375])\\ \hline
             Social Presence & (z=1.932, p=0.053, r=0.27, CI=[0.0,0.5]) & (z=0.0, p=1.0, r=0.0, CI=[-0.375,0.5])\\ \hline
             Embodiment & (z=0.316, p=0.752, r=0.04, CI=[-0.25,0.25])  & (z=-0.657, p=0.511, r=0.01, CI=[-0.333,0.167])\\ \hline
             \textbf{Scenario/ Dim}& \textbf{IPQ: Realness} &\\
             Place Illusion & (z=5.269, p<0.001, r=0.75, CI=[0.75,1.250]) &\\ \hline
             Involvement & (z=1.294, p=0.196, r=0.18, CI=[-0.125,0.375]) &\\ \hline
             Plausibility & (z=3.927, p<0.001, r=0.56, CI=[0.25,0.75]) &\\ \hline
             Social Presence & (z=2.623, p=0.009, r=0.37, CI=[0.0,0.5]) &\\ \hline
             Embodiment & (z=-0.098, p=0.922, r=0.01, CI=[-0.25,0.025])  & \\ \hline

        \end{tabular}
        \caption{Summary of the results of the Wilcoxon signed rank test for related samples for the presence dimensions of PQ and IPQ per condition as well as the corresponding 95\,\%-confidence interval.}
        \Description{Results of Wilcoxon signed-rank test per subscale of PQ and IPQ for positive and negative condition. From left to right are dimension, questionnaire, z-statistic, p-value, 95\% confidence interval, and effect size.}
        \label{tab:subscales_diff}
    \end{table}

\end{document}